\documentclass[a4paper,11pt]{article} 

\usepackage{jcappub} 
\usepackage[T1]{fontenc} 

\usepackage{newtxtext,newtxmath}
\usepackage{amsmath}
\usepackage{amssymb}
\usepackage{graphicx}
\graphicspath{{plots/}}
\usepackage{bm}
\usepackage[dvipsnames]{xcolor}
\usepackage{xspace}
\usepackage{ulem}
\usepackage{lipsum}
\usepackage{xfrac}
\usepackage{jcappub}
\usepackage{aas_macros}
\begin{document}

\title{Towards field-level likelihood for projected fields: \\Evolved projected fields from initial projected fields}

\author{Kevin Hong$^{1,3}$}
\author{Rugved Pund$^{2,3}$}
\author{An\v ze Slosar$^3$}

\affiliation{$^1$University of California, Los Angeles, Department of Physics \& Astronomy, Los Angeles, CA 90095, USA}
\affiliation{$^2$Physics and Astronomy Department, Stony Brook University, Stony Brook, NY 11794, USA}
\affiliation{$^3$Brookhaven National Laboratory, Physics Department, Upton, NY 11973, USA}

\abstract{
       The evolved cosmological matter density field is fully determined by the initial matter density field at fixed cosmological parameters.  However, the two-dimensional cosmological projected matter density field, relevant for weak-lensing and photometric galaxy studies, is fully determined by the initial projected matter density field only at the linear order. At non-linear order,  the entire volume of initial matter contributes.  We study a model for the evolved projected density field that is deterministic in the initial projected density fields and probabilistic in the effects of the remaining modes in the initial conditions. We write down predictions for the mean evolved projected field model using Lagrangian perturbation theory. We run a suite of small $N$-body simulations with fixed projected initial conditions and measure the statistical properties of the ensemble of evolved projected fields. Measurements and theory are in good agreement and show that the information on the initial projected fields is exponentially  suppressed on non-linear scales. 
       We implement this approach in a likelihood code  and use Hamiltonian Monte-Carlo sampling to show that, at fixed cosmology, initial projected fields can be reconstructed even in the presence of non-trivial mask features. We validate the reconstruction and its uncertainties on an ensemble of realizations.
}

\maketitle

\section{Introduction}\label{sec:intro}

Density fluctuations in the universe are one of the fundamental probes of cosmology. Their statistics and evolution can tell us about the basic cosmological parameters as well as test fundamental physical theories.  Fluctuations in the universe evolve from the primordial fluctuation seeded by inflation in the early universe. The cosmological initial conditions are generally believed to be close to a Gaussian random field. Under the effects of gravity, these initial fluctuations evolve into a rich non-Gaussian field that can still be described as a random field, whose statistical description is constrained by statistical homogeneity and isotropy.  Theories of structure formation cannot predict where the universe will be more or less dense, but they can predict an arbitrary summary statistics of the field, such as 2-point correlation function or a power spectrum.

For a given set of based cosmological parameters, the realization of the
primordial fluctuations uniquely determines the evolved matter density
fields\footnote{Strictly speaking, this is not true in the presence of baryons
and chaotic small scale behavior, but these effects can be neglected for the
current discussion.}. This can be done, for example, by running an $N$-body
simulation in the computer.  In that sense, the evolved density field can either
be considered a random field whose correlators depend on cosmological
parameters, or a deterministic field, which depends on all the cosmological
parameters and the initial density fluctuations.   This dichotomy has led to two
approaches in cosmological inference. The first approach relies on measuring
various summary statistics of the tracers of cosmological structure and then
fitting models to it -- for example measuring the galaxy cluster, weak-lensing
shear and their cross-correlation functions, also known as 3$\times$2 analysis
and then fitting those measurement with theoretical predictions.  An alternative
approach, still in its infancy, is to directly fit the observed over-density
fields as functions of not only cosmological parameters, but also the full
vector of initial conditions. This latter approach, known as the field-level
likelihood has a strong advantage that can, in principle, extract \emph{all}
information present in the evolved density fields \cite{arXiv:1203.3639,1806.11117,2103.04158,2509.09673,2501.16852,2507.05378,2504.15351}.

The field-level likelihood approaches usually suffers from the dimensionality curse, namely the complexity associated with the very large number of parameters needed to fully describe the initial conditions, which scale with the total volume that needs to be described.   The number of free parameters can therefore easily go into millions and evolving three-dimensional boxes in a numerically efficient manner can also be daunting. Starting with a two-dimensional field-level likelihood on cosmological quantities that are inherently two-dimensional, such as weak-lensing field and photometric galaxy clustering therefore sounds like an attractive stepping stone towards the full three-dimensional field-level analysis
\cite{2108.04825,2312.08934}. The number of parameters should, at face value, scale with area rather than volume and forward modelling a two-dimensional field sounds easier than performing the three-dimensional problem. Other routes to the non-Gaussian information in projected fields include machine-learning emulation of the fields \cite{arXiv:2402.07694,2304.06099} and higher-order statistics \cite{2409.07980,2408.16847}. Moreover, field-level likelihood can also naturally deal with systematic effects, such as very complex masks that appear in weak lensing: the mask is simply applied to the forward-modelled field \cite{2312.08934}, while in summary-statistics analyses it requires dedicated window-convolution modelling \cite{2202.11839}.

At linear order, the evolved projected density fields are uniquely determined by the initial projected density fields -- they are simply scaled by the growth factor. Unfortunately, at non-linear order, this is not true: the evolved projected density fields depends on the initial projected density fields and also all the other modes that affect its evolution through mode coupling. In this paper we will refer to those as bulk modes. Based on counting the available degrees of freedom it is obvious that the information about the bulk modes is very degenerate in the projected field, which makes the 2D field level likelihood no easier and perhaps even more difficult than fitting the full 3D field directly. 

Naively, one could bite the bullet and simply fit both the initial projected and bulk modes and use that in fitting the projected fields \cite{2108.04825}.  However, an alternative approach is to hybridise between field-level likelihood and summary statistics approach. Namely, the dominant contribution, which is the effect of the initial projected modes on the evolved projected modes is treated deterministically and the residual effects of the bulk modes is treated probabilisticaly in terms of translationally and rotationally invariant correlators. The way to to think about it is the following. Consider the set of all possible realization of (three-dimensional) initial conditions. For each initial condition there is a corresponding evolved matter field. The ensemble mean of both initial and evolved fields is zero.  Next consider as subset of initial conditions that produce a required initial projected conditions. Such set has a corresponding well-defined set of evolved density fields and a corresponding set of evolved projected fields. On large scales we know that the mean of these fields needs to follow the linear theory. What happens on weakly non-linear scales is the subject of this paper.

This paper is organized as follows. In Section \ref{sec:theory} we develop a theory of evolved projected fields using Lagrangian perturbation theory. We test this theory in the Section \ref{sec:simulations}. In Section \ref{sec:likelihood} we implement a likelihood and perform sampling to reconstruct the initial conditions. In the final section \ref{sec:discussion} we discuss how this can be applied to a real-world scenario and conclude. In this exploratory paper we focus on the dark matter only.

This paper has undergone significant revisions and bug fixes with respect to its earlier versions. These are summarized in Appendix \ref{app:changes}.

\newcommand{\vb}[1]{\mathbf{#1}}
\newcommand{\vk}{{\vb{k}}}
\newcommand{\vq}{{\vb{q}}}
\newcommand{\vx}{{\vb{x}}}
\newcommand{\vp}{\vb{p}}
\newcommand{\vpsi}{\vb{\Psi}}
\newcommand{\vkperp}{{\vb{k}_{\perp}}}
\newcommand{\kperp}{{k}_\perp}
\newcommand{\kpar}{{k_{\parallel}}}
\newcommand{\qpar}{{q_{\parallel}}}
\newcommand{\Pperp}{P_{\perp}}

\newcommand{\vxperp}{{\vb{x}_{\perp}}}
\newcommand{\xpar}{{x_{\parallel}}}
\newcommand{\hatP}{\hat{P}}

\section{Theory}\label{sec:theory}
\subsection{Preliminaries}

Given some three-dimensional over-density field $\delta(\vx)$ and its Fourier transform $\delta(\vk)$\footnote{From argument to $\delta$ it is clear whether we mean a real-space or Fourier space.}, we define the projection operator as
\begin{align}
    \hat{P} \left(\delta(\mathbf{x})\right) \equiv \int d x_\parallel \, W(x_\parallel) \delta(x_\parallel, \mathbf{x}_\perp),
\end{align}
where $W(x_\parallel)$ is the radial window function. The window function has units of inverse length and is normalized so that $\int W(\xpar) d\xpar = 1$. In Fourier space
\begin{align}
\hat{P}{\delta}(\mathbf{k}) = \int \frac{dk'_{\parallel}}{2\pi} \ W_k(k'_{\parallel}) \delta(k'_{\parallel}, \mathbf{k_{\perp}}) 
\end{align}
and $W_k$ is the Fourier transform of the window function.  Normalization requires $W_k(0)=1$. $\hat{P}\delta(\vkperp)$ is related to $\hat{P}\delta(\vx_\perp)$ using the usual 2D Fourier transform.

The field $\delta$ is a standard cosmological over-density field satisfying 
\begin{eqnarray}
    \left<\delta(\mathbf{k}) \right>_G&=&0\\
    \left<\delta(\mathbf{k})\delta(\mathbf{k}')\right>_G&=&(2\pi)^3\delta^D(\mathbf{k}+\mathbf{k}')P(k)\\
    \left<(\hatP \delta)(\kperp) \right>_G&=&0\\
    \left<(\hatP \delta)(\kperp) (\hatP\delta)(\kperp')\right>_G&=& (2\pi)^2 \delta^D(\kperp+\kperp') P_{\rm 2D} (\kperp)
\end{eqnarray}
with 
\begin{equation}
    P_{\rm 2D}(\kperp) = \int W^2(\kpar) P(\kperp,\kpar) d\kpar 
\end{equation}
Here we use the subscript $G$ to denote a global average over all possible cosmologies. Now, of all possible realizations of $\delta$, we want to pick a subset of realizations that have a fixed projected modes, i.e. those for which $\hatP \delta(\kperp) = d(\kperp)$. There are infinitely many realization of $\delta$ that satisfy this condition. 

It is well known fact that a Gaussian distribution that is conditioned on the value of some values of the field (or linear combinations thereof) remains a Gaussian distribution with a different mean and covariance.  The ensemble of initial conditions with identical projected modes can be constructed by
\begin{align} \label{eq:init-cond}
        \Delta(\mathbf{k}) \equiv \delta(\mathbf{k}) + 2\pi \delta^D(\kpar)  \left(d(\kperp)- \hat{P}{\delta}(\mathbf{k})\right).
\end{align}
It is easy to show that $\hatP \Delta(\vk)=d(\kperp)$ and since $\hatP \Delta(\vk)$ has no delta dependence it means that $\left<\hatP \Delta(\vk)\right> = d(\kperp)$ and its variance ${\rm Var}\, \hatP \Delta(\vk) = 0$. This is the field that we want.
For a projection over the full periodic box, which is the case realized in our simulations, the projected modes are exactly the $\kpar=0$ modes, these are statistically independent of all the other modes and Equation \ref{eq:init-cond} is the exact conditional ensemble. For a general window $W$ the constraint is a linear combination of modes with different $\kpar$ and the exact conditional ensemble requires a covariance-weighted correction spread over all those modes. Equation \ref{eq:init-cond} then still defines an ensemble with the required projected modes, but one in which the entire correction is carried by the $\kpar=0$ modes. The results of this section apply to this ensemble and are exact conditional expectations only in the full-box case.

In standard cosmology, the initial conditions are completely unconstrained and the evolved field, after averaging over all possible realizations of the initial field has a zero mean and some finite power spectrum (and higher order correlators). Starting with $\Delta$ as initial conditions and evolving it, the evolved field will not be a zero mean, since it retains the memory of fixed projected mode in the initial conditions. In the next section we will calculate this in the context of Lagrangian Perturbation theory.
\subsection{Lagrangian Perturbation Theory}

Using the continuity equation in terms of the Lagrangian coordinates $\mathbf{x}(t)=\mathbf{q}+\mathbf{\Psi}_\mathbf{q}(t)$, the time evolution of the initial conditions in the Zeldovich approximation can be written as,
\begin{align} \label{eq:zeldovich}
	\Delta(\mathbf{k},t) = -(2\pi)^{3}\delta^{D}(\mathbf{k}) + \int d^{3}\mathbf{q} \ \exp\left(- i \mathbf{k} \cdot [\mathbf{q} + \mathbf{\Psi}_{\mathbf{q}}(t)]\right)
\end{align}
where the displacement field $\mathbf{\Psi}(\mathbf{q},t)$ is linearly proportional to the initial conditions $\Delta_0$

\begin{align}
	\mathbf{\Psi}(\mathbf{q}, t) = D(t) \int_{\vk} e^{i \mathbf{k} \cdot \mathbf{q}} \frac{i \mathbf{k}}{k^{2}} \Delta_{0}(\mathbf{k})
\end{align}
Here and in what follows, we  use a notational shorthand $\int_{\vk} \equiv \int  \frac{d^{3}\mathbf{k}}{(2\pi)^{3}}$. Substituting Eq.\ref{eq:init-cond} and taking the ensemble mean gives (dropping $\delta^D(\vk)$ terms), 
\begin{align}
	\left< \Delta(\mathbf{k},t)\right>
	 & = \int_{\mathbf{q}} \exp\left[ -i \mathbf{k} \cdot \left(\mathbf{q} + D(t)  \mathbf{\Psi}_{0}^{(d)}(\mathbf{q})\right) \right]
	\ \left< \exp\left[ -i \mathbf{k} \cdot D(t) \mathbf{\Psi}_{0}^{(\delta)}(\mathbf{q}) \right] \right>
\end{align}
where $\mathbf{\Psi}_{0}^{(d)}$ is the fixed component and $\mathbf{\Psi}_{0}^{(\delta)}$ is the component of the initial conditions that depend on $\delta$.
\begin{align*}
\mathbf{\Psi}_{0}^{(\delta)}(\mathbf{q}) &= \int \frac{d^{3}k}{(2\pi)^{3}} \, e^{i \mathbf{k} \cdot \mathbf{q}} \frac{i \mathbf{k}}{k^{2}} \left[ \delta(\mathbf{k}) - 2\pi\delta(k_{\parallel}) \hat{P} \delta(\mathbf{k}_{\perp}) \right] \\
\mathbf{\Psi}_{0}^{d}(\mathbf{q}) &= \int \frac{d^{2}k}{(2\pi)^{2}} e^{i \mathbf{k}_{\perp} \cdot \mathbf{q}_{\perp}} \frac{i \mathbf{k}_{\perp}}{k_{\perp}^{2}} d(k_\perp)
\end{align*}

The characteristic function of the initial displacement field $\mathbf{\Psi}^{(\delta)}_{0}$ can be simplified\footnote{For a Gaussian random field $\varphi(x)$, $\log \langle e^{it\varphi(x)} \rangle = it\langle \varphi(x) \rangle - t^{2}\langle \varphi(x) \varphi(x) \rangle^{2}/2$ using the cumulant expansion theorem. 
Also note that $(\mathbf{1} - \hat{P})\delta_\mathbf{k}$ is a linear transformation of the Gaussian random field $\delta(\mathbf{k})$ and hence also a Gaussian random field.},
\begin{align}
	\left< \exp \left[ -i \mathbf{k} \cdot D(t) \mathbf{\Psi}_{0}^{(\delta)}(\mathbf{q}) \right] \right>
	 & = \exp\left[ - \frac{1}{2} D^{2} \left< \mathbf{k} \cdot\mathbf{\Psi}^{(\delta)}(\mathbf{q}) \ \mathbf{k} \cdot\mathbf{\Psi}^{(\delta)}(\mathbf{q}) \right> \right]
\end{align}

Simplifying the variance,
\begin{align}
        \left< \mathbf{\Psi}^{(\delta)}(\mathbf{q}) \mathbf{\Psi}^{(\delta)}(\mathbf{q}) \right>
        = & \ \int_{\mathbf{k'}} \frac{\mathbf{k'} \mathbf{k'}}{k'^{4}} P(\mathbf{k'}) \\ \nonumber
        & + \int_{\mathbf{k'}} \frac{(\vkperp',0)}{{\kperp'}^{2}}
        \frac{({\vkperp'},0)}{{k'}_{\perp}^{2}} |W(k'_\parallel)|^2 P(\mathbf{k'}) \\ \nonumber
        & - 2 \int_{\mathbf{k'}} e^{i k'_{\parallel} \cdot q_{\parallel}} \frac{\mathbf{k'}}{k'^{2}}
        \frac{({\vkperp'},0)}{{\kperp'}^{2}} W(k'_{\parallel}) P(\mathbf{k'}) \\ \nonumber
\end{align}
and therefore we find
\begin{equation}
k_{i} k_{j} \left< \psi^{(\delta)}_{i}(\mathbf{q}) \psi^{(\delta)}_{j}(\mathbf{q}) \right>
        = k^2 \Sigma_Z^2 + \kperp^2 \Sigma^2_{W^2} - \kperp^2 \Sigma^2_{W}(\qpar),
\end{equation}
where
\begin{align}
    \Sigma_{Z}^2 =&  \frac{1}{6\pi^2} \int dk' \, P(k') \\
    \Sigma{^2}_{W^2} =&  \frac{1}{4\pi^2} \int_0^\infty d\kpar' W(k'_{\parallel})^{2} L_2(\kpar') \\
    _{\rm full}\Sigma{^2}_{W} (\qpar) =& \frac{1}{2\pi^2} \int_0^\infty d \kpar' \cos(\kpar' \qpar) W(\kpar') L_1(\kpar') \label{eq:cos}\\
    L_2(\kpar') =& \int_0^\infty d \kperp' \frac{P(k')}{{\kperp'}} \\
    L_1(\kpar') =& \int_0^\infty  d \kperp' \frac{P(k')\kperp'}{{k'^2}} = \int_\kpar^\infty dk' \frac{P(k')}{k'} 
\end{align}

We see that $\Sigma_W$ depends on $\qpar$. This is because our projection operator breaks parallel translational symmetry. Under transformation $\xpar \rightarrow \xpar + \Delta \xpar$, the $\hatP(\delta(\vx))$ will change.  $W(\kpar)$ has support only at low $\kpar$, so we expand the cosine inside expression for $\Sigma^2_W$ to write:
\begin{align}
    _{\rm full}\Sigma{^2}_{W} (\qpar) =& \Sigma^2_W - \frac{1}{2}\qpar^2 \Sigma^2_{(2)W} + \ldots  \\
    \Sigma{^2}_{W} (\qpar) =& \frac{1}{2\pi^2} \int_0^\infty d \kpar'  W(\kpar') L_2(\kpar) \\
    \Sigma{^2}_{(2)W} (\qpar) =& \frac{1}{2\pi^2} \int_0^\infty d \kpar' \kpar^2 W(\kpar') L_2(\kpar) 
\end{align}

Since $W(\kpar)$ has support at low $\kpar$ this should be a convergent series with $\Sigma^2_{(2)W}$ small compared to $\Sigma^2_W$ as long as windows are large compared to non-linear scale. Simplifying and putting it all together,
\begin{align}
        \left\langle \Delta(\mathbf{k},t) \right\rangle
&= e^{ -\frac{1}{2}D^{2} \left( \kpar^2 \Sigma^2_\mathrm{Z} + k_{\perp}^{2} \left(\Sigma^2_\mathrm{Z}+\Sigma^2_{W^{2}}- \Sigma^2_{W}\right)\right) } \int_{q_{\parallel}} e^{ -\frac{1}{2}D^{2} k_{\perp}^{2} \qpar^2 \Sigma^2_{(2)W} } \int_{\mathbf{q}_{\perp}}  e^{ -i \mathbf{k}_{\perp} \cdot \left(\mathbf{q}_{\perp} + D(t)  \mathbf{\Psi}_{0}^{(d)}(\mathbf{q}_{\perp})\right) }
\end{align}

We see that the $\qpar$ integral just changes the overall normalization and the same holds for the projection operator. Since we know that a theory needs to 
reproduce the linear theory on the largest scales, we find that

\begin{equation}
\Delta(\Delta_0,t) = \hat P  \left\langle \Delta(\mathbf{k},t) \right\rangle = e^{ -\frac{1}{2} D^2 \kperp^{2}\Sigma^2} \mathcal{Z}(\Delta_0,t),
\label{eq:key}
\end{equation}
where 
\begin{equation}
    \Sigma^2 = \Sigma^2_\mathrm{Z}-\Sigma^2_{W}+\Sigma^2_{W^{2}}
\end{equation}
is the total suppression and 
\begin{equation}
\mathcal{Z}(\Delta_0,t) = e^{ -i \mathbf{k}_{\perp} \cdot \left(\mathbf{q}_{\perp} + D(t)  \mathbf{\Psi}_{0}^{(d)}(\mathbf{q}_{\perp})\right) }
\end{equation}
is a linear field evolved in a Zeldovich approximation to time $t$.  This is the key result of this paper. This recipe can be summarized as follows: the evolved projected 2D field is a two-dimensional evolution of the initial field, multiplied by a Zeldovich-like suppression that we know from the standard Lagrangian theory. The first term in this suppression is exactly the same as the suppression of BAO wiggles, with an important distinction that it applies to the \emph{one}-point function rather than the \emph{two}-point function (hence the extra factor of a half). The interpretation, however is the same: at the same fixed initial projected modes, the different realization of the bulk modes will push the resulting structures in the different directions -- when averaging over those different directions a smearing appears that reduces the power at high k. The correction terms $\Sigma^2_{W}$ and $\Sigma^2_{W^2}$ reduce the overall damping taking into account that some modes are evolved explicitly and therefore do not contribute to damping. Note also that we use a sign convention that makes all $\Sigma^2$ quantities positive. 

The shape of the suppression is Gaussian, but only to the leading order in which we expand the cosine in the equation \ref{eq:cos}. We also see that the damping factor multiplies a Zeldovich-evolved 2D field. At this point it is tempting to replace the $\mathcal{Z}$ operator with a generally evolved non-linear field to "re-sum" the corrections that would presumably appear if the calculations was led to a higher order. While not theoretically robust, this is a common swindle.

\section{Comparison with Simulations.}\label{sec:simulations}
\newcommand{\gadget}{{\texttt{GADGET-4}}\xspace}

\begin{figure}
    \includegraphics[width=\linewidth]{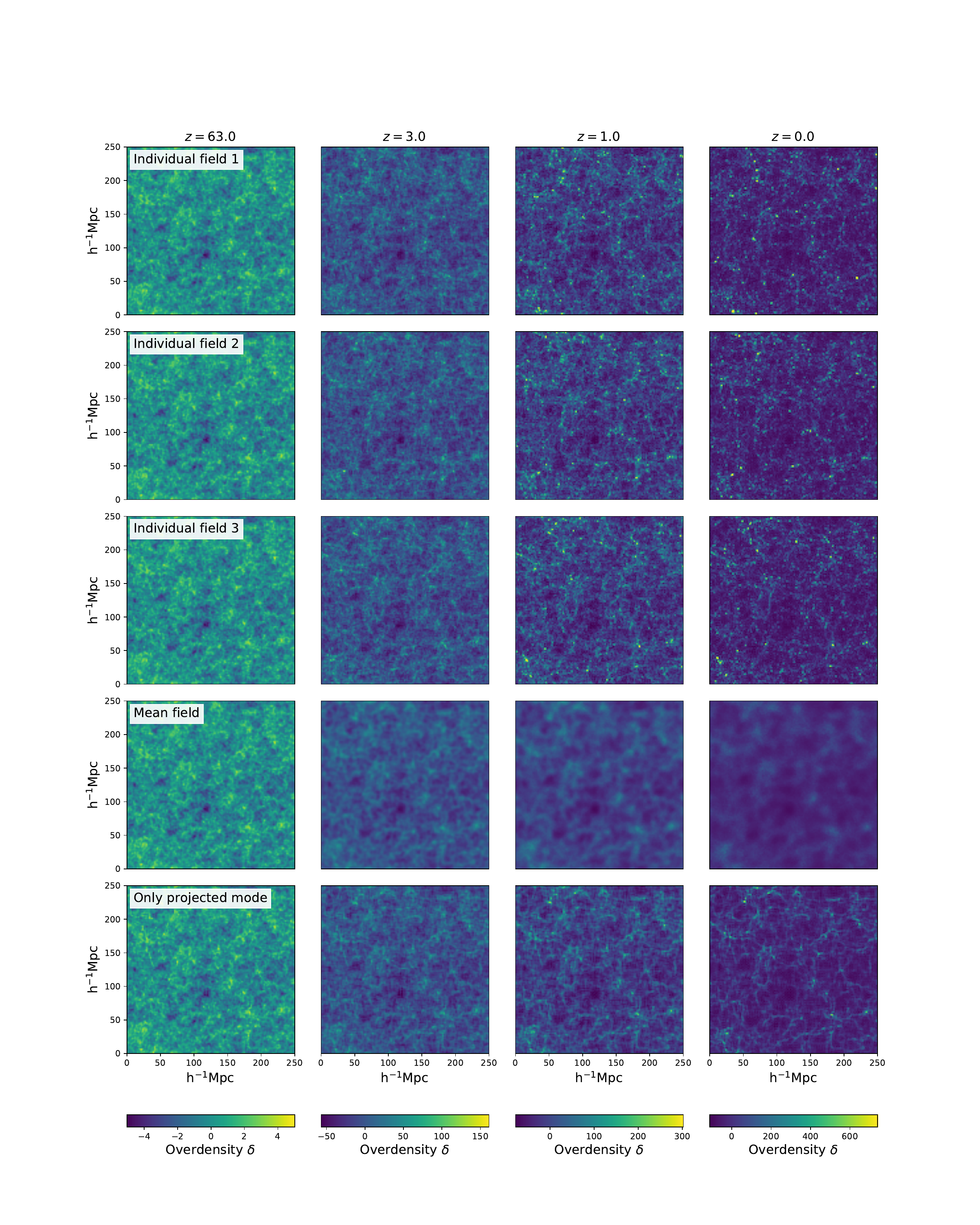}
    \caption{This figure shows the result of our simulations. In each panel we show the three-dimensional density field projected along the $z$ axis on the $x-y$ plane. Top three rows show three randomly chosen  realization out of the 100 we have run, the second-to-last row corresponds to the field-level mean of realizations and the final plot is for the projected-mode only. Columns from left to right correspond to decreasing redshifts as labeled on top. The fourth row (mean field) is the ensemble average over the 100 realizations at fixed projected modes, i.e.\ the field-level mean $\langle\Delta\rangle$ predicted by Equation \ref{eq:key}; its small-scale structure is suppressed because averaging over the bulk modes smears the displaced structures. The bottom row (projected-mode only) is the deterministic two-dimensional evolution $\mathcal{Z}(\Delta_0,t)$ of the projected modes alone, evolved with no bulk-mode coupling. Note that the plotted dynamic range is adapted at every redshift, but is uniform across the plots (see the bottom color-scales). See text for the discussion. }
    \label{fig:projplots}
\end{figure}

\subsection{Approach}
In this section we will test the result presented against theory simulations. To this end, we run a suite of 100 small simulations that, crucially, had the same projected initial conditions. This simulation suite is not meant to be competitive for comparison with data, but to provide a sufficiently accurate test-bed for the theoretical predictions. We chose $N=128^3$ dark-matter-only particles in a periodic cube with sides of comoving length $L=250\ h^{-1} \rm Mpc$. This box is sufficiently large that the largest scale modes remain in the linear regime to $z=0$, while maintaining sufficient resolution to faithfully represent transition to non-linear regime.  We used \gadget to perform simulations \citep{arXiv:2010.03567,arXiv:astro-ph/0003162,2103.04158,arXiv:1401.5466}.

Initial conditions were generated using Gadget's internal IC generator, which has been modified to allow for fixed projected fields. This was achieved by employing two pseudo-random number generators with two seeds. The first seed, held fixed, was used to generate modes $(k_x, k_y, k_z)$ with $k_z=0$, while the second seed, different for each of the 100 simulation, was used to generate the remaining IC modes with $k_z\neq 0$. Initial conditions are generated based on second-order Lagrangian perturbation theory \cite{arXiv:astro-ph/0606505} at an initial redshift of $z_{\textrm{init}} = 63$. The initial power spectrum has been generated using Efstathiou approximation to the linear dark matter power spectrum \cite{1992MNRAS.258P...1E}.

We run an additional simulation, which we refer to as ``Projected Modes Only'' (PMO), in which  the projected field was initialized as above, but all the remaining modes were set to zero. This simulation is in effect an evolved 2D cosmological field (in a 3D cosmological background). 

The cosmological parameters used to evolve simulation box were fixed to default \gadget values: $\Omega_0 = 0.308$, $\Omega_\Lambda = 0.692$, $\Omega_b = 0.0482$, $h = 0.678$, $n_s = 1.0$, $\sigma_8 = 0.9$, where symbols have their conventional meaning in cosmology.  Every simulation box was evolved to redshift 0.

For each output snapshot file at a specific redshift, we interpolate the particle positions onto a density mesh using a cloud-in-cell (CIC) interpolation scheme. We can then project those fields along the $z$ direction to get the two-dimensional projected fields.

\subsection{Results}

The over-density of these projected fields is plotted in the Figure \ref{fig:projplots}. This plot illustrates most of the effects relevant for this discussion, so it is worth spending some time on. 
At the very high-redshift (left-most column), the universe is linear and therefore the projected modes evolve independently of the rest of the box. The rest of the box is Gaussian distributed and adds to exactly zero.  Miniscule differences that can be observed between boxes at this initial redshift can be attributed the the 2LPT that has been used to evolve the boxes to this redshift.  As we move towards lower redshift, the upper three panels show the non-linear structure formation. Note that this a field projection, rather than a slice which is plotted more often, therefore the web-like structure is somewhat less present, but one can clearly see dark-matter halos in projection. Staring at the three individual realizations independently we see that while the overall structure is coherent, the exact positions at which the haloes in projection appear varies from realization to realization.  When we compare this with projected-only mode plotted in the bottom, we see that the latter contains fewer isolated peaks since those correspond to truly three-dimensional concentration of density, but that the web structure is more pronounced. Finally, we see that the second from the bottom panel, the field-level average is heavily suppressed on small scales. This is exactly as expected given Equation \ref{eq:key}. The effect of three-dimensional modes it to push small scale structure in one-direction in one realization and a different direction in a different realization resulting in an overall smearing of small scale structure.  In Appendix  \ref{app:x} we show the same figure but for the projection along the $x$-axis.

\begin{figure}
    \includegraphics[width=\linewidth]{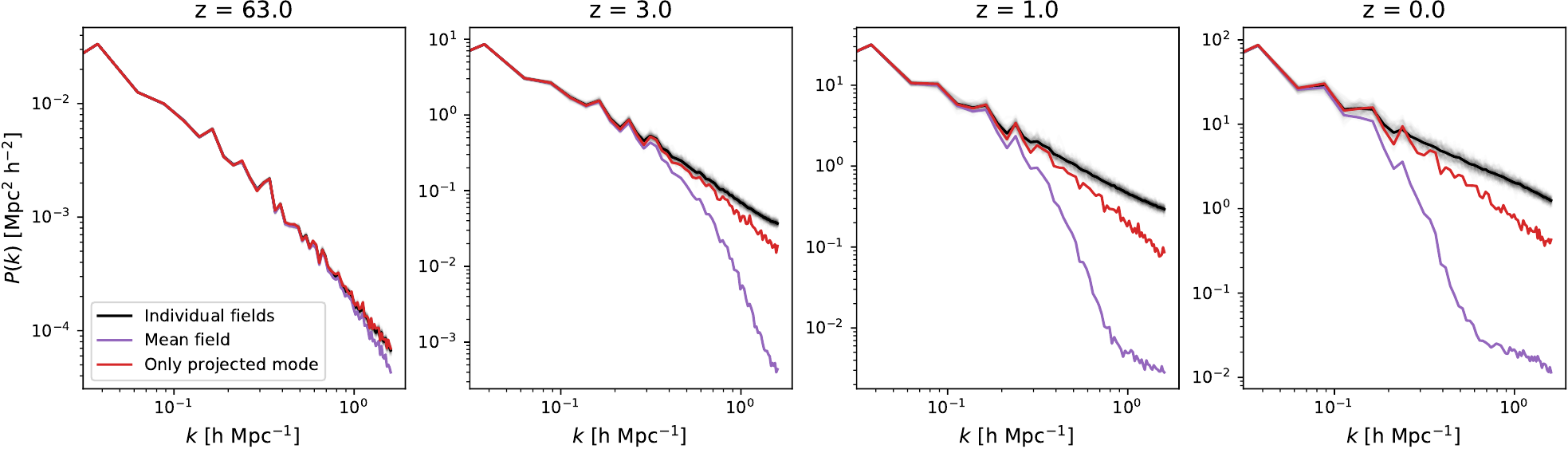}
    \caption{Two-dimensional power spectra for 100 simulations. Black line corresponds to the mean power spectra of 100 realizations (at fixed initial projected modes; top 3 rows in Figure~\ref{fig:projplots}). The red line shows the only-projected mode realization (bottom row in Figure~\ref{fig:projplots}). The purple line shows power spectra of the mean field (second-to-last row in Figure~\ref{fig:projplots}).}\label{fig:pspec}
\end{figure}
In Figure \ref{fig:pspec} we show resulting two-dimensional power spectra. At early times, all three power spectra track the same linear prediction as expected. Since the mean power spectra are mean over 100 realizations at fixed projected modes they are very low-noise compared to to other spectra. Since a single realization is an unbiased measurement of the standard projected power spectrum (since it is evolved from bona fide initial conditions), this is very close to the expected standard projected power spectrum. The red lines corresponds to the power spectrum of projected-modes only. The fact that red is somewhat suppressed with respect to black is a result of missing power from bulk modes scattering into projected modes (i.e. contribution of two modes with wave-numbers $(\kperp,+\kpar)$ and $(\kperp,-\kpar)$.). The purple line shows the very strong suppression discussed above.  The correct way to understand the purple line is that the system is forgetting its initial projected state in projected. The effect of coupling of bulk and projected modes means that information about the initial projected modes gets propagated into bulk modes and vice-versa: when only projected modes are available, the information is effectively lost.

To make a quantitative comparison, we first calculate the value of $\Sigma^2$. We set $W(\kpar) = {\rm sinc}(\kpar L/2)$, corresponding to a top-hat window of size $L$ and evaluated the integrals numerically for the cosmology and initial power spectrum corresponding to our simulation suite. Results can be found in Table \ref{tab:num}. As expected, we find total $\Sigma^2$ to be similar in magnitude and somewhat smaller than the purely Zeldovich $\Sigma^2_Z$. 

\begin{table}
    \centering
\begin{tabular}{c|ccc|c}
    L / (Mpc/h) & $\Sigma^2_\mathrm{Z}$ / (Mpc/h)$^2$& $\Sigma^2_{W^2}$ / (Mpc/h)$^2$ & $\Sigma^2_W$ / (Mpc/h)$^2$& $\Sigma^2 = \Sigma^2_\mathrm{Z}-\Sigma^2_{W} +\Sigma^2_{W^{2}}$ / (Mpc/h)$^2$\\
    \hline
    100 & 35.4 & 45.1 & 51.5 & 29.0  \\
150 & 35.4 & 30.1 & 40.7 & 24.9  \\
\textbf{250} & \textbf{35.4} & \textbf{18.0} & \textbf{28.4} & \textbf{25.1}  \\
500 & 35.4 & 9.0 & 16.1 & 28.4  \\
1000 & 35.4 & 4.5 & 8.5 & 31.4  \\
2000 & 35.4 & 2.3 & 4.4 & 33.3  \\
\end{tabular}
\caption{Values of suppression factors evaluated for simulation cosmology for various values of $L$ at z=0. The value relevant to the simulation box size $L=250$Mpc/h is emphasized in bold. Values of $\Sigma^2$ at other redshift scale as the square of the linear growth factor $D(z)$ (normalized to $D(z{=}0)=1$), and therefore \emph{decrease} with increasing redshift.}
\label{tab:num}
\end{table}

We now consider the ratio between the mean projected and projected-only modes
\begin{equation}
     \frac{P_{me}(\kperp)}{P_{ee}(\kperp)} = \exp\left(-\frac{1}{2}\kperp^2 \Sigma^2\right), \label{eq:plot}
\end{equation}
where index $m$ corresponds to the mean evolved projected field and index $e$ to the non-linear evolved projected field. The evolution of the former is traced by Equation \ref{eq:key}, while the latter is simply $\mathcal{Z}(\Delta_0,t)$ giving the exponential suppression as all that remains.

\begin{figure}
    \includegraphics[width=\linewidth]{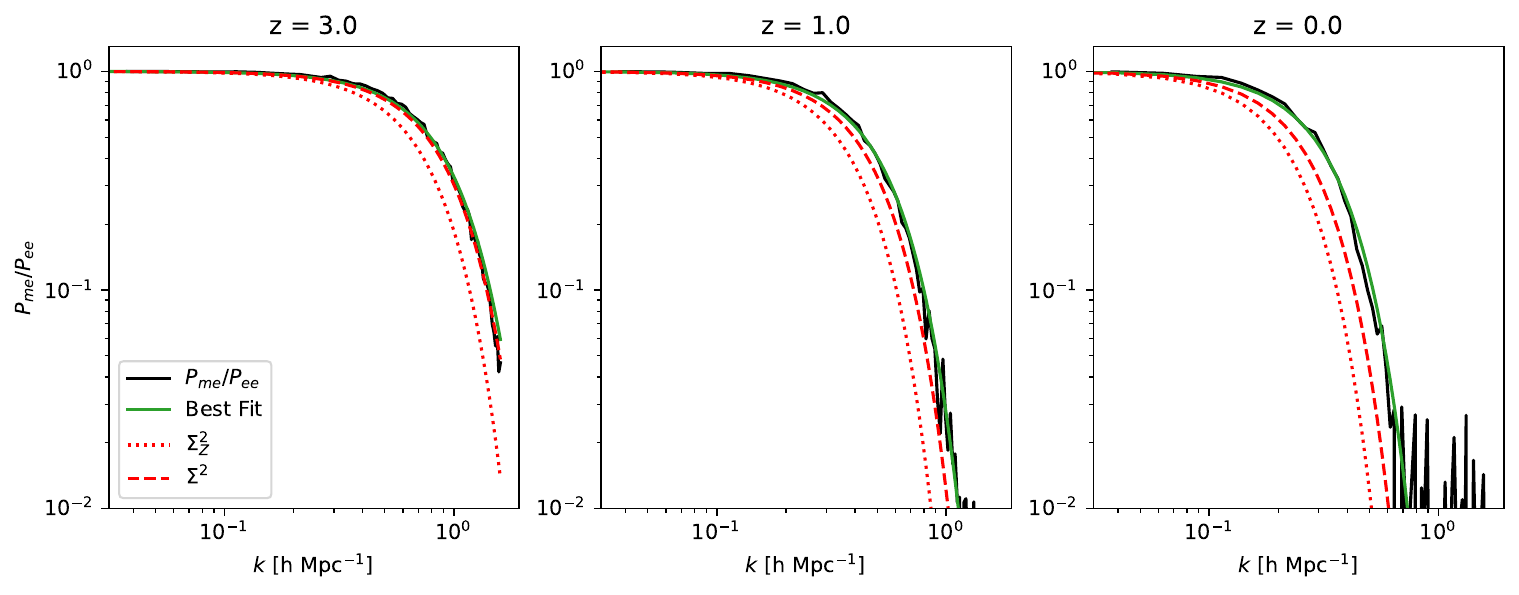}
    \caption{The quantity of Equation \ref{eq:plot} measured at three different redshifts. We plot a theoretical suppression $\Sigma^2$ (red dashed) as well as suppression expected from Zeldovich only $\Sigma_Z^2$ (red dotted) and a theoretical fit $\Sigma_{\rm fit}^2$ (green) against the measured values (black solid line). The fit assumes the single-parameter Gaussian form of Equation \ref{eq:plot}. }
    \label{fig:suppr}
\end{figure}
 
In Figure \ref{fig:suppr} we plot simulation results together with theoretical expectations. We find that phenomenologically the model works very well, the Gaussian suppression is well supported by the data down to the non-linear scales we can measure. We also see that our correction to Zeldovich term is significant and improves agreement with theory. We find that after taking into account the growth factor, the suppression is predicted to be $\Sigma^2= 25.1,\,9.3,\,2.5$\,Mpc/$h^2$ at redshifts $z=0,\,1,\,3$ respectively. Fitting the Gaussian form of Equation \ref{eq:plot} to the measured ratio $P_{me}/P_{ee}$ of the simulation suite gives $\Sigma^2\approx 17.2,\,7.2,\,2.2$\,Mpc/$h^2$ at the same redshifts. The measured values lie $\sim$10--30\% below the theoretical prediction, with the discrepancy growing towards lower redshift where higher-order corrections become important. Interestingly, the measured $\Sigma^2$ values are \emph{lower}, implying \emph{less} suppression, i.e the linear modes track the evolved modes to higher redshift than our theory implies.

\begin{figure}
    \centering
    \includegraphics[width=\linewidth]{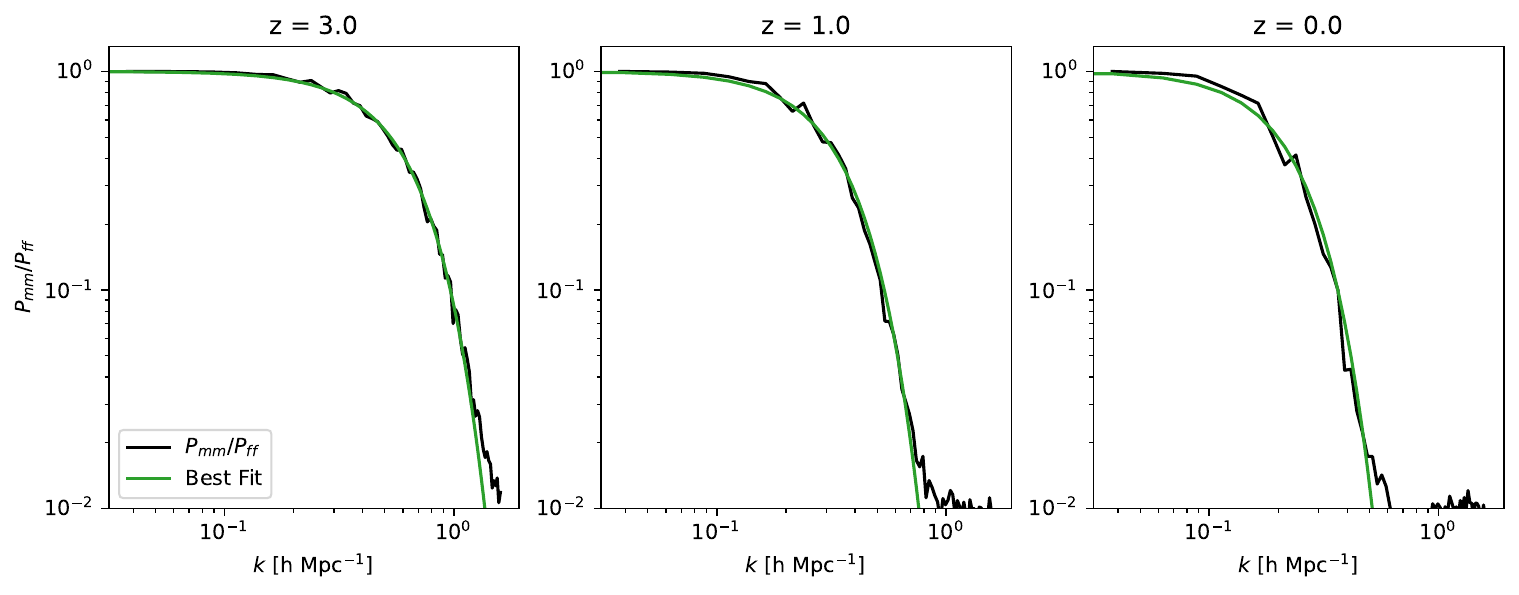}
    \caption{As Figure \ref{fig:suppr}, but for the suppression measured relative to the \emph{full} non-linear power spectrum, $P_{mm}/P_{ff}=\exp(-\kperp^2\Sigma^2)$ (Equation \ref{eq:plot_full}). Black is the measured ratio and green the Gaussian best fit. This is the signal-power fraction used by the likelihood of Section \ref{sec:likelihood}.}
    \label{fig:suppr_full}
\end{figure}

It is also instructive to measure the suppression relative to the \emph{full} non-linear power spectrum (the black line in Figure \ref{fig:pspec}) rather than to the projected-mode-only power $P_{ee}$. This is the quantity the likelihood of Section \ref{sec:likelihood} effectively uses: there the residual variance is anchored to the full power $P_{ff}$, so the relevant object is the signal-power fraction
\begin{equation}
    \frac{P_{mm}(\kperp)}{P_{ff}(\kperp)} = \exp\left(-\kperp^2 \Sigma^2\right),
    \label{eq:plot_full}
\end{equation}
where $m$ is the mean (evolve-then-project) field and $f$ a full non-linearly evolved realization. Note that, since the mean field is itself the ensemble average of the full fields, $\langle P_{mf}\rangle = P_{mm}$ exactly, so $P_{mf}/P_{ff}$ is simply a noisier estimator of the same ratio; and because $P_{mm}$ is an auto-spectrum, Equation \ref{eq:plot_full} carries the full $\kperp^2$ rather than the factor $\tfrac12$ of the cross-spectrum in Equation \ref{eq:plot}, while measuring the same $\Sigma^2$. This estimator has no clean theoretical derivation, but the Figure \ref{fig:suppr_full} shows that as a phenomenological model it works well. Since the full power is above non-linearly evolved projected power, we need more suppression and therefore $\Sigma^2$ values creep up. Results are summarized in Table \ref{tab:suppr}. We find that suppression against full spectrum tracks the theoretical predictions even better, but absent theory we believe this is a fluke.

\begin{table}
\centering
\begin{tabular}{c|ccc}
$z$ & predicted $25.1\,D^2(z)$ & measured $P_{me}/P_{ee}$ & full-power $P_{mm}/P_{ff}$ \\
\hline
0 & 25.1 & 17.2 & 17.4 \\
1 &  9.3 &  7.2 &  8.0 \\
3 &  2.5 &  2.2 &  2.5 \\
\end{tabular}
\caption{The suppression $\Sigma^2$ (in $(\mathrm{Mpc}/h)^2$) at three redshifts, determined three ways: the theory prediction $25.1\,D^2(z)$ (Table \ref{tab:num} scaled by the growth factor); the value measured from the cross-spectrum ratio $P_{me}/P_{ee}=\exp(-\tfrac12\kperp^2\Sigma^2)$ (Equation \ref{eq:plot}); and the value measured relative to the full non-linear power as the signal-power fraction $P_{mm}/P_{ff}=\exp(-\kperp^2\Sigma^2)$ (Equation \ref{eq:plot_full}). All three are the same $\Sigma^2$ entering the mean-field damping $\exp(-\tfrac12 D^2\kperp^2\Sigma^2)$. The last column is the empirical, robust estimator adopted by the likelihood of Section \ref{sec:likelihood}.}
\label{tab:suppr}
\end{table}

\section{Likelihood and Sampling}
\label{sec:likelihood}

The result of Section \ref{sec:theory}, Equation \ref{eq:key}, predicts only the
\emph{ensemble mean} evolved projected field. A field-level likelihood requires the full
conditional distribution $P(\Delta_{\rm obs} | \Delta_0)$, where $\Delta_{\rm obs}$ is the observed field and $\Delta_0$ are the initial condition. In this
section we write down a concrete likelihood built around Equation \ref{eq:key}, implement
it as a differentiable forward model, and use it to demonstrate that we can recover the initial fields on large scales \emph{beyond} simply linear modes.

\subsection{The likelihood and its design choices}
\label{sec:like:def}
We model the observed projected field as the suppressed two-dimensional evolution of the
initial projected field plus a stochastic residual,
\begin{equation}
  \Delta_{\rm obs}(\kperp) = S(\kperp)\,\mathcal{Z}(\Delta_0,t)(\kperp) + n(\kperp),
  \qquad S(\kperp) = e^{-\frac12 D^2 \kperp^2 \Sigma^2},
  \label{eq:like:model}
\end{equation}
where $\mathcal{Z}(\Delta_0,t)$ is the deterministic 2D evolution of the initial projected
field (Eq.~\ref{eq:key}) and $n$ is the residual sourced by the unknown bulk modes.

The generative model is the following. The initial projected modes are Gaussian with the known linear projected power spectrum, $\Delta_0 \sim \mathcal{N}(0,P_0)$. Given $\Delta_0$, the observed field is Gaussian with mean $S\mathcal{Z}(\Delta_0,t)$ and a covariance that is diagonal in $\kperp$ with variance
\begin{equation}
  N(\kperp) = \big[1 - S^2(\kperp)\big]\,P_{\rm tot}(\kperp),
  \qquad S^2(\kperp) = e^{-D^2 \kperp^2 \Sigma^2},
  \label{eq:like:noise}
\end{equation}
where $P_{\rm tot}$ is the \emph{total} non-linear projected power spectrum (the black line in Figure \ref{fig:pspec}). This is the signal-power fraction of Equation \ref{eq:plot_full}: if a fraction $S^2$ of the total power is carried by the mean field, the residual carries the remaining $1-S^2$. We stress that, as discussed at the end of the previous section, this is a phenomenological model in which $\Sigma^2$ is defined with respect to the total power; it does not follow from the derivation of Section \ref{sec:theory}.

At fixed cosmology $P_0$ and $P_{\rm tot}$ are known functions. For a given $\Sigma^2$ the Bayes theorem then gives the posterior of the initial modes
\begin{equation}
  -\ln P(\Delta_0 | \Delta_{\rm obs}) = \frac12 \sum_{\kperp} \bigg[
   \frac{|\Delta_0(\kperp)|^2}{P_0(\kperp)}
  + \frac{|\Delta_{\rm obs}(\kperp) - S(\kperp)\,\mathcal{Z}(\Delta_0,t)(\kperp)|^2}{N(\kperp)}
  \bigg] + {\rm const}.
  \label{eq:like:logl}
\end{equation}
The first term is the Gaussian prior on the initial modes and the second the likelihood of the residual. For $P_0$ we use the linear power spectrum from which the simulation initial conditions were drawn. For $P_{\rm tot}$ we use the mean power spectrum of the simulation suite. To make it independent of the fields that we analyse, we measure it from the projections of the 100 simulations along the $x$ and $y$ axes: these are 200 projected fields with the same statistical properties as the $z$ projections, but with independent initial projected modes. In an analysis of real data $P_{\rm tot}$ would come from a model of the non-linear power spectrum and would vary with cosmological parameters together with $P_0$ and $\Sigma^2$.

Note that $P_{\rm tot}$ is an input and not a free parameter. The data cannot determine the amplitude of the residual separately from the signal on the scales where the signal dominates, and a posterior in which $P_{\rm tot}$ is left free is dominated by its prior (see Appendix \ref{app:changes}).

Modelling the residual in this way involves three approximations.  First, as Figure \ref{fig:projplots} shows, $n$ is clearly non-Gaussian: it contains precise positions of halos from the 3D field which simply cannot be predicted from the initial projected modes alone. For a model that is linear in its parameters, a Gaussian likelihood with the correct two-point function still gives unbiased estimates\footnote{In standard derivations of optimal quadratic estimator one starts with a likelihood that is strictly speaking true for Gaussian fields only, but the implied expression for power spectrum estimate is manifestly unbiased for arbitrarily non-Gaussian field \cite{astro-ph/9611174}.} and a Gaussian model was found to give unbiased results in the weak-lensing analysis of \cite{2312.08934}, where the dominant stochastic term is, however, the nearly Gaussian shape noise. Our forward model is non-linear in $\Delta_0$, so this protection is not guaranteed. Second, at fixed $\Delta_0$ the residual is not statistically homogeneous, since there is more of it where the projected density is high. In our suite the variance of the residual on the scales of interest ($\kperp<0.4\,h\,{\rm Mpc}^{-1}$) is $2.2$ times larger in the densest quarter of the area of the predicted field than in the least dense one and the map of this variance has a correlation coefficient of $0.8$ with the predicted density. The corresponding correlations between pairs of different Fourier modes are individually small, with an rms correlation coefficient of about $0.03$, but they are spread over many pairs. Its covariance is therefore not diagonal in $\kperp$: statistical isotropy makes it diagonal only after averaging over $\Delta_0$ and it is this averaged covariance that we use. Third, $(1-S^2)P_{\rm tot}$ is only an approximation to the actual power of the residual about the implemented prediction, which we check in the next subsection. We are therefore trading optimality for simplicity. Whether the resulting posterior is unbiased and whether its width is meaningful are empirical questions, which we answer with an ensemble of realizations in Section \ref{sec:like:ensemble}.

We make the following design
choices for this exploratory demonstration:
\begin{itemize}
  \item We \emph{fix} the cosmology and with it the initial-mode prior $P_0$ and the total power $P_{\rm tot}$ throughout. A complete cosmological analysis
        would instead sample $\Omega_m$ and $\sigma_8$, which control both, as well as $\Sigma^2$. In Section \ref{sec:like:fixed} we in addition fix $\Delta_0$ to its true value and determine $\Sigma^2$ and the forward model from the data. In Sections \ref{sec:like:sampling} and \ref{sec:like:ensemble} we fix these to their best-fit values and sample
        the primordial projected modes $\Delta_0$.
  \item We focus on $z=1$ as a representative, mildly non-linear case.
  \item We consider three choices for the forward evolution $\mathcal{Z}$ (the 2D LPT
        displacement equations we implement are collected in Appendix \ref{app:lpt}): Eulerian linear
        (EL; $\mathcal{Z} = D(t)\,\Delta_0$, i.e.\ no displacement), first-order Lagrangian
        (1LPT, Zeldovich), and second-order Lagrangian (2LPT).
  \item The free parameters are the initial modes with $\kperp$ up to the Nyquist frequency of the $128^2$ grid, which are the modes present in the initial conditions of the simulations.
  \item A survey mask $W$ multiplies both the data and the predicted mean field $S\mathcal{Z}$ in real space before the transformation to Fourier space, where Equation \ref{eq:like:logl} is evaluated with the same $N$. In other words, the residual is set to zero in the masked area. This is the usual $f_{\rm sky}$ approximation of pseudo-spectrum analyses, where $f_{\rm sky}$ is the unmasked fraction of the area: a mode of the masked residual has variance $f_{\rm sky}N$, but only a fraction $f_{\rm sky}$ of the modes are independent, and the two factors cancel in the $\chi^2$ of Equation \ref{eq:like:logl}. For a white residual the result is exact, since by Parseval's theorem the sum over modes is then the sum over the observed pixels. For our coloured residual it is an approximation, which neglects the correlations of the residual across the mask boundaries; we test it by repeating the analysis without the mask.
\end{itemize}

The likelihood has been implemented in the \texttt{PyTorch} framework in python. This makes it run natively on a GPU and be automatically differentiable. 
The code to reproduce all results in this paper is publicly available at \url{https://github.com/slosar/ProjectedFields}.

\subsection{Fixed initial conditions: which evolution, and what $\Sigma^2$?}
\label{sec:like:fixed}
\begin{figure}
    \centering
    \includegraphics[width=0.6\linewidth]{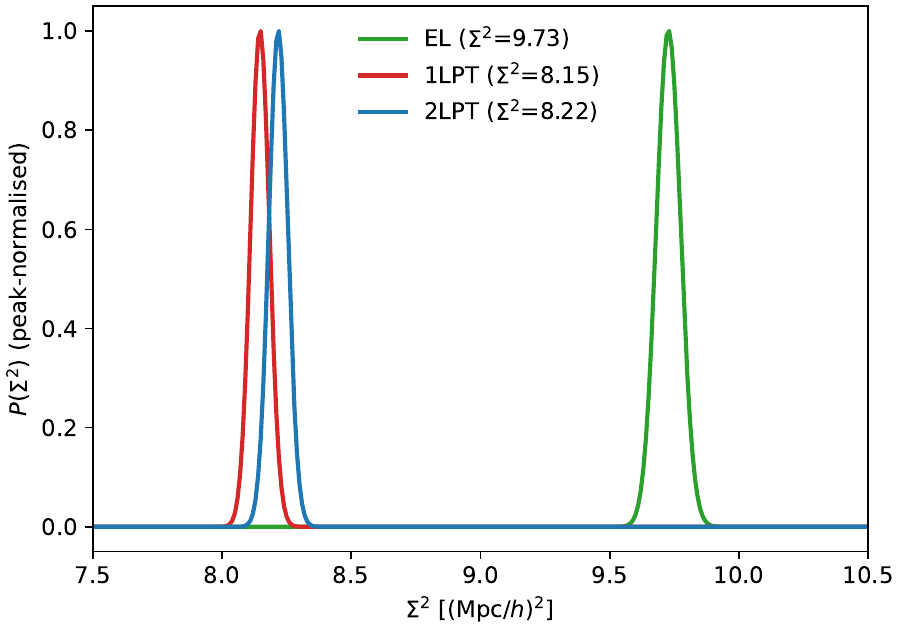}
    \caption{Combined constraint on $\Sigma^2$ from the 100-realization suite at $z=1$ with
    the initial field fixed to truth, for the three forward models (peak-normalized,
    $P(\Sigma^2) \propto \exp\sum_i \ln\mathcal{L}_i$). 1LPT and 2LPT agree and recover
    $\Sigma^2 \approx 8.2$; Eulerian linear prefers a larger value.}
    \label{fig:like_sigma2}
\end{figure}

We first fix $\Delta_0$ to its true value (the projected-mode-only initial field) and ask
what value of $\Sigma^2$ the data prefer and which forward model fits best. Summing
$\ln\mathcal{L}$ (which here includes the normalisation term $-\frac12\sum\ln N$, since $N$ depends on $\Sigma^2$, and uses the mask of the next subsection) over all 100 realizations at $z=1$ as a function of $\Sigma^2$ gives the
combined constraint shown in Figure \ref{fig:like_sigma2}. The 1LPT and 2LPT cases peak at
$\Sigma^2 = 8.15\pm0.03$ and $8.22\pm0.03\ (\mathrm{Mpc}/h)^2$ respectively, consistent with the
previous section. The Eulerian-linear case peaks
markedly higher, $\Sigma^2 = 9.73\pm0.04$: lacking the Zeldovich displacement, EL retains too
much coherent small-scale power and requires extra damping to match the data.

\begin{figure}
    \centering
    \includegraphics[width=\linewidth]{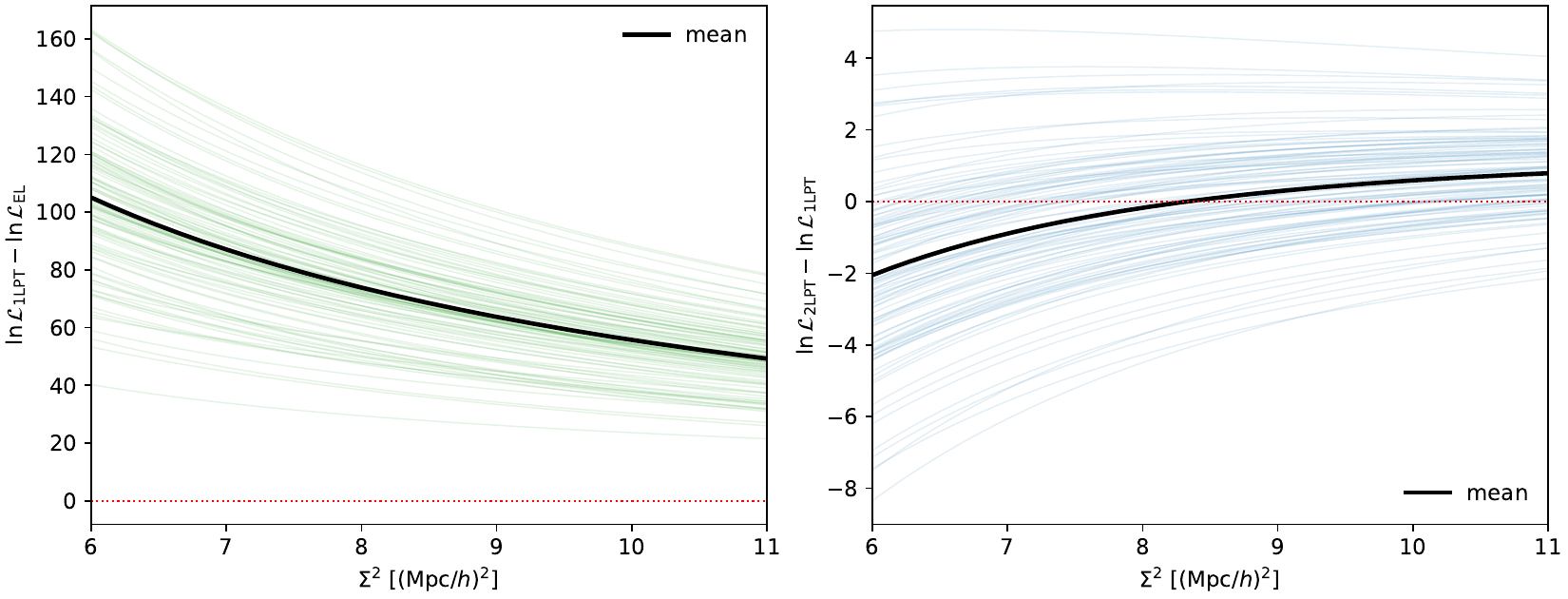}
    \caption{Per-realization log-likelihood differences with the initial field fixed to truth
    (thin lines; thick line is the mean over realizations). \emph{Left}: 1LPT versus Eulerian
    linear --- the displacement is decisive. \emph{Right}: 2LPT versus 1LPT --- no significant difference.}
    \label{fig:like_orders}
\end{figure}

Figure \ref{fig:like_orders} compares the forward models directly, as the per-realization
difference in log-likelihood. Adding the Zeldovich displacement (1LPT over EL) improves the
fit enormously --- by $72$ in $\ln\mathcal{L}$ on average and by at least $29$ for \emph{every} realization --- showing
that the displacement, not merely the growth of the amplitude, is what matters. Going to
second order (2LPT over 1LPT) makes no measurable difference: at $\Sigma^2=8.2$ the mean difference is $\Delta\ln\mathcal{L}=-0.1$ with a scatter of $1.7$ between realizations and 2LPT is the better fit for 49 of the 100.

\begin{figure}
    \centering
    \includegraphics[width=\linewidth]{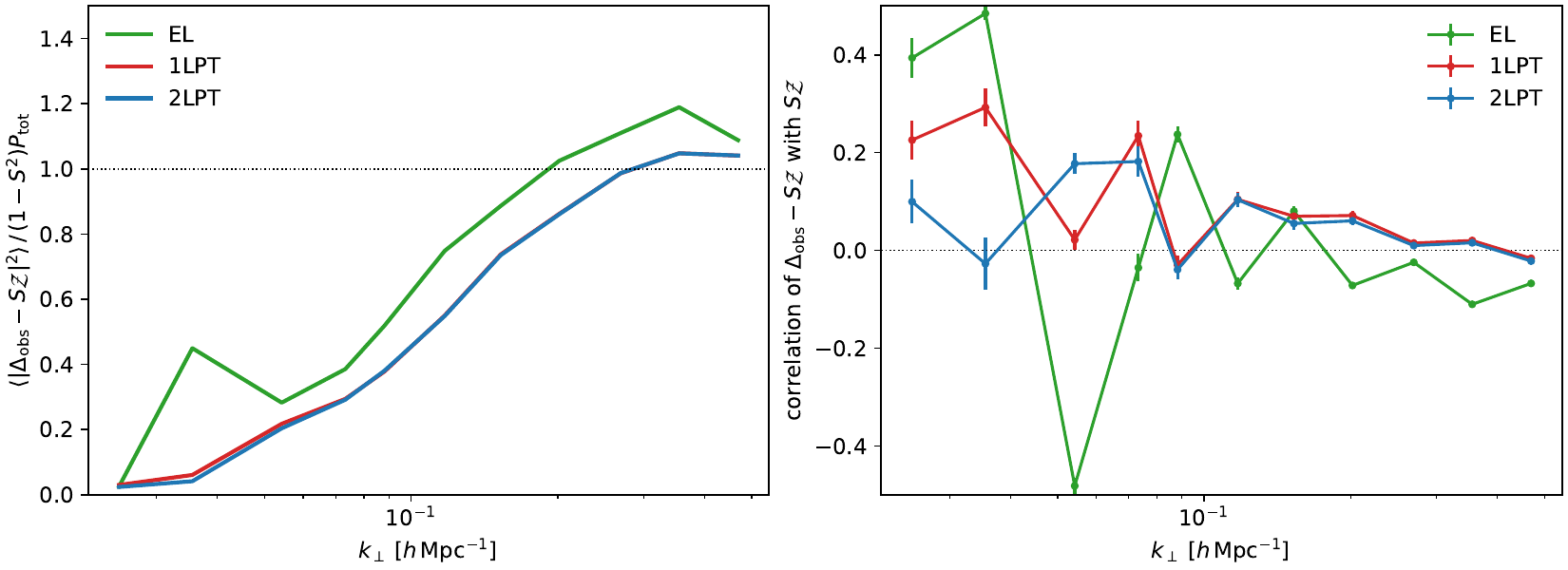}
    \caption{Checks of the residual model about the implemented prediction, with the initial field fixed to truth and averaged over the 100 realizations. \emph{Left}: power of the residual $\Delta_{\rm obs}-S\mathcal{Z}$ relative to the model variance $(1-S^2)P_{\rm tot}$. \emph{Right}: correlation coefficient between the residual and the prediction $S\mathcal{Z}$, with the error of the mean over realizations.}
    \label{fig:like_resid}
\end{figure}

The same setting allows us to check the two assumptions that the model makes about the residual, both taken about the implemented prediction $S\mathcal{Z}(\Delta_0)$ rather than about the true ensemble mean. The results are shown in Figure \ref{fig:like_resid}. First, the residual should be uncorrelated with the prediction. For both 1LPT and 2LPT the regression coefficient of the suite-mean field on the prediction is within $7\%$ of unity in every bin with $\kperp<0.5\,h\,{\rm Mpc}^{-1}$ and the correlation coefficient between the residual and the prediction does not exceed $0.1$ for $\kperp>0.08\,h\,{\rm Mpc}^{-1}$. It is larger on the largest scales (up to $0.3$), but there the residual itself is tiny. There is a small but significant positive correlation of $0.07$--$0.10$ at $\kperp\simeq 0.1$--$0.2\,h\,{\rm Mpc}^{-1}$, i.e.\ the prediction is slightly too low in amplitude there. Second, the power of the residual should be $(1-S^2)P_{\rm tot}$. This holds to within $5\%$ for $\kperp>0.25\,h\,{\rm Mpc}^{-1}$, but on larger scales the model \emph{overestimates} the residual power: by $15$--$35\%$ at $\kperp\simeq0.15$--$0.2\,h\,{\rm Mpc}^{-1}$, by a factor of two to three at $\kperp\simeq 0.1\,h\,{\rm Mpc}^{-1}$ and by more than an order of magnitude at the fundamental mode of the box. The likelihood is therefore conservative on large scales, where the reconstruction is in any case nearly perfect. Replacing $(1-S^2)P_{\rm tot}$ with a residual spectrum calibrated on simulations would be a straightforward improvement, which we do not pursue here.

Since the two LPT orders describe the data equally well and, as we show below, give indistinguishable reconstructions, we adopt the simpler 1LPT as our fiducial forward model and fix $\Sigma^2=8.2\ (\mathrm{Mpc}/h)^2$ in what follows.

\subsection{Sampling the initial field from masked data}
\label{sec:like:sampling}
\begin{figure}
    \centering
    \includegraphics[width=0.45\linewidth]{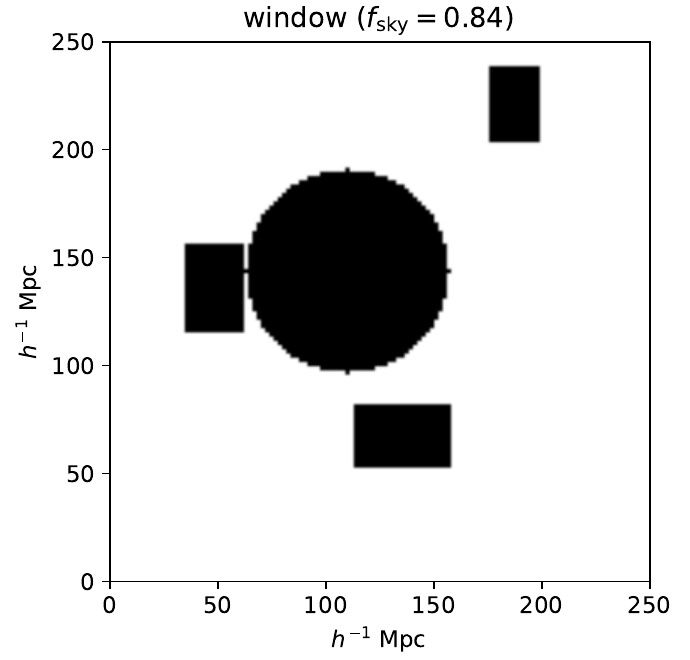}
    \caption{The survey window used in the sampling demonstration; black regions
    ($\approx 16\%$ of the area) are masked.}
    \label{fig:like_mask}
\end{figure}

We now infer $\Delta_0$ from a single realization, sampling the posterior of Equation \ref{eq:like:logl} (at fixed $\Sigma^2 = 8.2$, 1LPT
forward model) with Hamiltonian Monte Carlo. The initial field is parametrised by a real white-noise field $\xi$ of unit variance, with $\Delta_0(\kperp)=\sqrt{P_0(\kperp)}\,\xi(\kperp)$, so that the prior is a unit Gaussian and the sampler is preconditioned with the prior covariance. Every iteration is a trajectory of 50 leapfrog steps, with a step size that is adapted during burn-in to an acceptance rate of $75\%$ and then frozen. To mimic the masking of a realistic survey we multiply both the data
and the prediction by a window that masks $\approx 16\%$ of the area (Figure \ref{fig:like_mask}). All results in this subsection are for realization 88.

We run 8 independent chains, each started from a different random draw from the prior, with 300 burn-in iterations followed by 2000 iterations that we keep. The mean acceptance rate is $0.74$. The effective sample size of the amplitudes of the largest-scale modes and of the log-posterior is between 400 and 900 per chain, i.e.\ there is one effectively independent sample per two to five iterations. The chains agree with each other: the Gelman--Rubin statistic of these amplitudes is $\hat R<1.001$ and the posterior means of the individual chains differ by $2\%$ of the posterior standard deviation. The numbers for the 2LPT forward model are the same.
Because we are sampling a relatively small 2D field, sampling is fast: the eight chains together take about four minutes on a consumer grade GPU.

\begin{figure}
    \centering
    \includegraphics[width=\linewidth]{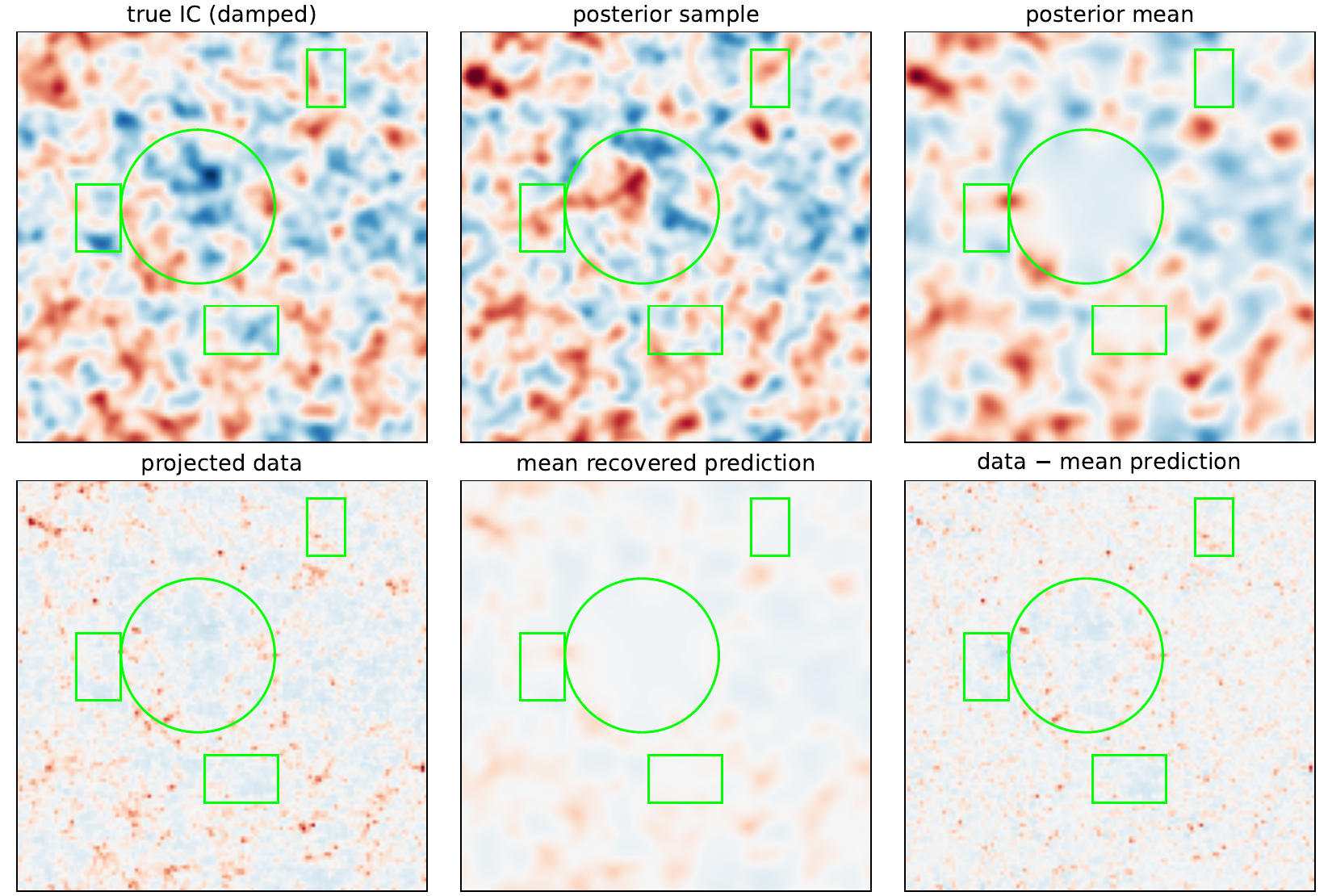}
    \caption{Initial-field recovery for realization 88 (1LPT, $\Sigma^2=8.2$), all damped by
    $S$. \emph{Top}: true initial field, one posterior sample, and the posterior mean.
    \emph{Bottom}: projected data, mean recovered prediction, and their difference. Mask
    boundaries are outlined in green.}
    \label{fig:like_recovery}
\end{figure}

Figure \ref{fig:like_recovery} shows the recovered initial field (top row) and the
corresponding data-space comparison (bottom row), all low-pass filtered by $S$ to suppress
the prior-dominated small scales. The posterior mean is a smooth, Wiener-like reconstruction
that matches the (filtered) truth on the scales that survive the suppression. The method
``infills'' across the mask boundaries, using the surrounding data and the prior, but it
cannot recover genuine structure deep inside the masked regions, where the posterior mean goes to zero.

\begin{figure}
    \centering
    \includegraphics[width=\linewidth]{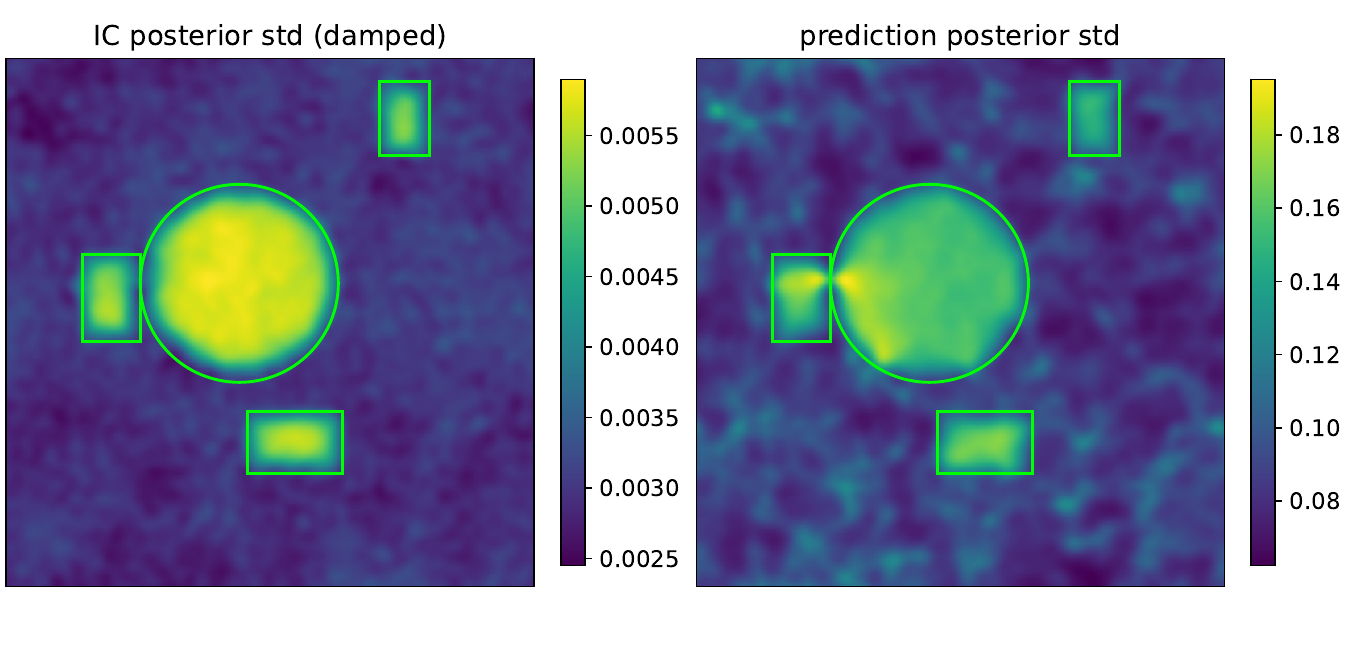}
    \caption{Posterior standard deviation across samples of the (damped) initial field
    \emph{(left)} and of the prediction \emph{(right)}, for realization 88. The variance
    rises inside the masked regions (outlined), where the data provide no constraint.}
    \label{fig:like_std}
\end{figure}

This is quantified by the posterior standard deviation (Figure \ref{fig:like_std}), which
rises sharply inside the mask --- by a factor of $1.7$ on average --- reflecting
the absence of a data constraint there. Each individual posterior sample places
statistically plausible but ultimately fictitious structure inside the holes; it is precisely
this realization-to-realization scatter of the unconstrained interior that drives the
increased variance.

\begin{figure}
    \centering
    \includegraphics[width=0.6\linewidth]{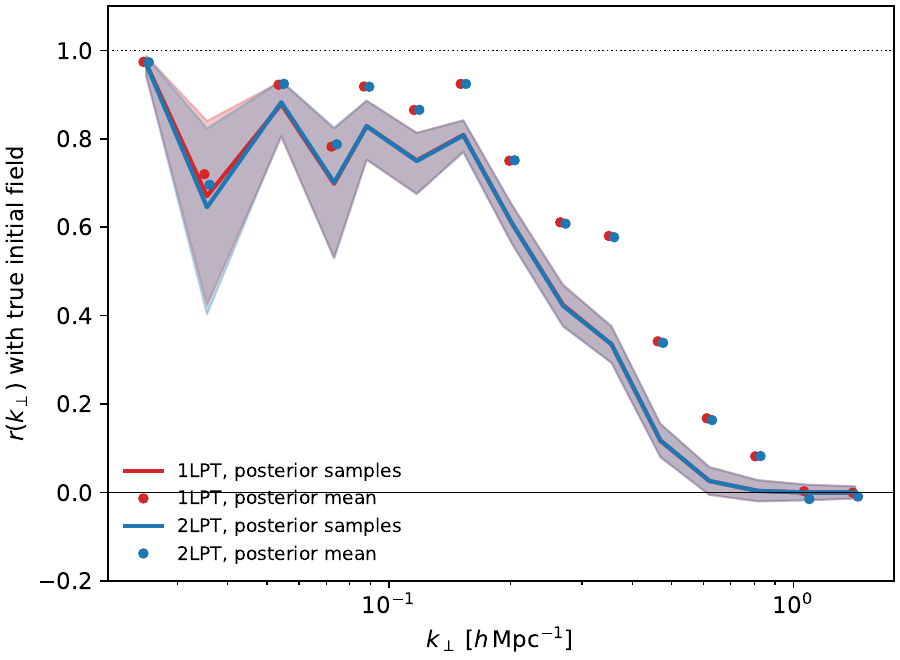}
    \caption{Cross-correlation of the recovered initial field with the truth for realization 88, comparing the 1LPT and 2LPT forward models. Lines and bands show the median and the central
    $68\%$ interval of the posterior samples and points the posterior mean. The two models are indistinguishable.}
    \label{fig:like_xcorr}
\end{figure}

Finally, Figure \ref{fig:like_xcorr} shows the cross-correlation coefficient between the
recovered and the true initial field, $r(\kperp) = P_{\rm fit\times true}/\sqrt{P_{\rm
fit}P_{\rm true}}$, for both 1LPT and 2LPT. It is close to unity on large scales, drops to about $0.6$ at $k_\perp \simeq 0.3\,h\,{\rm Mpc}^{-1}$ and is close to zero by $k_\perp \sim 0.6$--$0.8\,h\,{\rm Mpc}^{-1}$, beyond which the
information has been erased by the suppression. The posterior mean is better correlated with the truth than the individual samples, since the samples contain the posterior scatter. The dips at $\kperp\simeq0.04$ and $0.07\,h\,{\rm Mpc}^{-1}$ are a property of this particular $\Delta_0$, whose power in these two bins, which contain 4 and 8 modes, happens to be five times lower than the mean; the signal-to-noise ratio of these modes is correspondingly low. The two forward models give indistinguishable reconstructions.

\subsection{Tests on an ensemble of realizations}
\label{sec:like:ensemble}
\begin{figure}
    \centering
    \includegraphics[width=\linewidth]{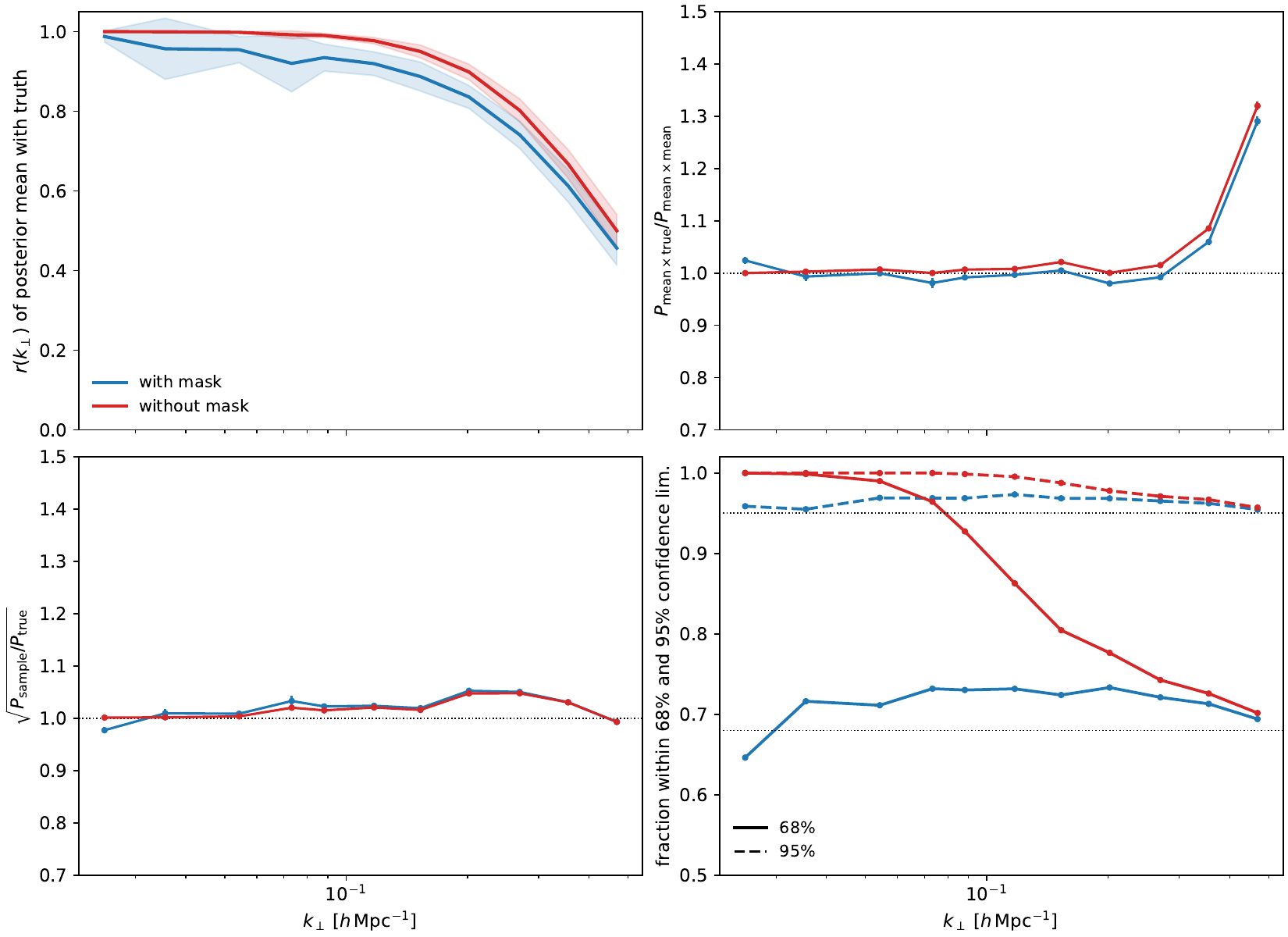}
    \caption{Ensemble tests of the reconstruction for 200 fields with independent $\Delta_0$, with (blue) and without (red) the mask. \emph{Top left}: cross-correlation coefficient of the posterior mean with the true initial field; lines are the mean over fields and bands the rms scatter between fields. \emph{Top right}: amplitude calibration of the posterior mean, which is unity for an unbiased reconstruction. \emph{Bottom left}: ratio of the amplitudes of the posterior samples and of the true field. \emph{Bottom right}: fraction of Fourier modes whose true value lies within the $68\%$ (solid) and $95\%$ (dashed) confidence limits of the posterior; dotted lines are the expected values. Error bars are bootstrap errors over fields.}
    \label{fig:like_ensemble}
\end{figure}

A single realization cannot tell us whether the reconstruction is unbiased or whether the posterior width is meaningful, and $r(\kperp)$ is by construction insensitive to the amplitude of the reconstruction. These are properties of the method on average over $\Delta_0$ and have to be tested on an ensemble of fields with different initial projected modes. The $z$ projections of our 100 simulations all share the $\Delta_0$ of the previous subsection and are therefore not suitable: the part of the reconstruction error that depends on $\Delta_0$, such as what is hidden by the mask, is the same in all of them. Instead we project the same 100 simulations along the $x$ and the $y$ axis. The initial projected modes of these 200 fields are different in every field\footnote{With the exception of the modes with $k_z=0$, which are fixed in all simulations by construction. These are two modes in every $|\kperp|$ shell, or less than $1\%$ of the modes.}, and the true initial projected field of each is known from the initial conditions of the corresponding simulation. We can therefore test the method on 200 different $\Delta_0$ without any new simulations. For these fields $P_{\rm tot}$ is measured leaving out the simulation to which the field belongs. Every field is analysed with the mask of Figure \ref{fig:like_mask} and without a mask, using one chain with 200 burn-in and 500 kept iterations (the median effective sample size is 190). The results are summarised in Figure \ref{fig:like_ensemble} and Table \ref{tab:coverage}.

\paragraph{Cross-correlation.} With the mask, the cross-correlation of the posterior mean with the truth is $0.99$ on the largest scales, $0.92$ at $\kperp\simeq0.1$, $0.84$ at $\kperp\simeq0.2$ and $0.61$ at $\kperp\simeq0.35\,h\,{\rm Mpc}^{-1}$. The scatter between fields is $0.03$--$0.04$ (up to $0.08$ in the bins with few modes). Without the mask the corresponding values are $1.00$, $0.98$, $0.90$ and $0.67$; the difference is expected, since $r$ is evaluated over the full area, of which $16\%$ is not observed.

\paragraph{Amplitude.} The transfer function $P_{\rm mean\times true}/P_{\rm true\times true}$ between the posterior mean and the truth is not the right test of bias: a posterior mean is shrunk towards zero wherever the data are not informative, so this ratio is below one by design (it equals $r^2$ for a calibrated posterior). The posterior mean is unbiased if the truth scatters symmetrically around it, i.e.\ if the regression coefficient of the truth on the posterior mean, $P_{\rm mean\times true}/P_{\rm mean\times mean}$, is unity. This coefficient is within $2.5\%$ of unity for all $\kperp<0.3\,h\,{\rm Mpc}^{-1}$, both with and without the mask, with statistical errors of $0.5$--$1\%$. It rises to $1.06$--$1.09$ at $\kperp\simeq0.35$ and to $1.3$ at $\kperp\simeq0.47\,h\,{\rm Mpc}^{-1}$, i.e.\ on these scales the posterior mean underestimates the amplitude of the modes. Little information is left there ($r\simeq0.6$ and $0.45$) and the error is on the conservative side. We attribute it to the forward model, since on these scales the correlation between $S\mathcal{Z}$ and the true mean field drops to $0.95$--$0.90$. As a second test, the power spectrum of the posterior samples should equal that of the true field. The ratio of amplitudes is between $0.98$ and $1.05$ at all scales, with an excess of $2$--$5\%$ at $\kperp\simeq0.1$--$0.4\,h\,{\rm Mpc}^{-1}$.

\paragraph{Confidence limits.} If the posterior width is meaningful, the true value should lie within the $68\%$ and $95\%$ confidence limits of the posterior in $68\%$ and $95\%$ of the cases. We test this in Fourier space, mode by mode (bottom right panel of Figure \ref{fig:like_ensemble}), and in real space, pixel by pixel (Table \ref{tab:coverage}). With the mask, the true value of a Fourier mode lies within the $68\%$ ($95\%$) limits for $69$--$73\%$ ($96$--$97\%$) of the modes at all $\kperp$, except for the fundamental modes of the box where the fractions are $65\%$ and $96\%$, which is consistent with the expected values given the small number of these modes. Without the mask the limits are conservative on large scales, where the fractions approach $100\%$, and approach the expected values at $\kperp>0.25\,h\,{\rm Mpc}^{-1}$. This is what one expects from Figure \ref{fig:like_resid}, since the model overestimates the residual power on large scales. In real space, for the damped initial field of Figures \ref{fig:like_recovery} and \ref{fig:like_std}, the limits are slightly conservative in the observed area and as expected inside the masked regions. The posterior standard deviation of Figure \ref{fig:like_std} and the bands of Figure \ref{fig:like_xcorr} can therefore be read as uncertainties in the usual sense.

\begin{table}
\centering
\begin{tabular}{l|cc|cc}
 & \multicolumn{2}{c|}{$68\%$ limits} & \multicolumn{2}{c}{$95\%$ limits} \\
 & observed & masked & observed & masked \\
\hline
with mask    & 0.76 & 0.68 & 0.97 & 0.95 \\
without mask & 0.76 & --   & 0.97 & --   \\
\end{tabular}
\caption{Real-space counterpart of the bottom right panel of Figure \ref{fig:like_ensemble}: the fraction of pixels of the damped initial field for which the truth lies within the $68\%$ and $95\%$ confidence limits of the posterior, separately for the observed area and for the area inside the mask, for the 200 fields with independent $\Delta_0$.}
\label{tab:coverage}
\end{table}

\paragraph{Mask and forward model.} The amplitude calibration and the confidence limits are essentially the same with and without the mask, so our treatment of the mask is adequate at this level. Repeating the masked analysis with the 2LPT forward model changes none of the above: the mean $r(\kperp)$ differs from that of 1LPT by less than $0.01$ in every bin and the calibration and the confidence limits agree within the errors. Finally, replacing the suite-mean $P_{\rm tot}$ with the binned power spectrum of each analysed field itself (tested on the 100 $x$ projections) also leaves the results unchanged.

\section{Discussion \& Conclusions}
\label{sec:discussion}

In this work we have calculated the relation between initial and evolved projected dark matter fields. At the linear level, the relation is deterministic as the field simply scale with the growth factor. At the non-linear level, the relation ceases to be deterministic, since it depends on the unknown configuration of the bulk modes. Still, an ensemble average over those unknown modes can be calculated. We have shown that this equals to the 2D evolution of the projected modes multiplied by an exponential suppression factor. The latter is similar in physics to the suppression factors that damps the BAO fluctuations, but applied to a field (rather than power spectrum) and contains corrections due to the projected modes that are not contributing to it. This suppression factor encodes the information loss due to the presence of bulk modes and shows that gains from the field level likelihood for 2D fields are going to be considerably less than those for the full 3D field.

To demonstrate the usefulness of this approach we have implemented a differentiable likelihood in the \texttt{PyTorch} framework. Likelihood sampling is extremely fast even on a consumer GPU and therefore analysis of real data is realistic with limited resources. What we have demonstrated is the reconstruction of the initial projected field at fixed cosmology, i.e.\ at fixed linear power $P_0$, suppression $\Sigma^2$ and total non-linear power $P_{\rm tot}$. Our numerical experiments have shown that the likelihood scheme can successfully sample from the initial conditions and recover the initial field (where recoverable) even in the presence of a non-trivial window function. Tests on 200 fields with independent initial conditions show that the amplitude of the reconstruction is unbiased at the level of $2.5\%$ for $\kperp<0.3\,h\,{\rm Mpc}^{-1}$ and that the $68\%$ and $95\%$ confidence limits of the posterior contain the truth as often as they should or slightly more often, despite the fact that the residual is modelled as a Gaussian field with a diagonal covariance. The exception are the smallest scales that still carry some information and where the reconstruction is biased low. The entire model improvement comes from 1LPT with 2LPT not bringing a statistically significant improvement. Importantly, the non-liner LPT1 model produces very significantly better fit than first order Eulerian description, showing that we are not simply copying modes from high redshift to low and dividing them by the growth factor, but that non-linear mapping is crucial in extracting information.

The step from here to cosmological inference is not trivial. Once $\Omega_m$ and $\sigma_8$ are varied, $P_0$, $\Sigma^2$ and $P_{\rm tot}$ all have to be predicted as functions of cosmology. The total power can then no longer be treated as a nuisance quantity and any error in the residual model, such as the overestimate of the residual power on large scales that we find for the simple form $(1-S^2)P_{\rm tot}$, will propagate into the cosmological parameters. We expect that most of the remaining difficulty lies there. We have also not quantified how much information the field-level approach gains with respect to the standard two-point analysis of the same projected field, the angular power spectrum $C_\ell$. Given the exponential loss of information on non-linear scales, the gain cannot be taken for granted and has to be established by an explicit comparison, which we leave for future work.

In our likelihood, we have modeled the residual field as Gaussian. This is essentially the same as the real non-linear field with large scale modes filtered out. This field therefore has its own power spectrum that can be predicted and used to infer cosmological parameters.  To get even more information and a better reconstruction of primordial modes, it can be explicitly modeled using approaches like normalizing flows (see e.g. \cite{2202.05282}).  Alternatively, a potential model would be a full 3D sampling of modes, which would naturally explain most of the additional structure in the data and therefore recover even more information. Of course, this is a highly degenerate problem where we are modeling $N^2$ measurement with $N^3$ free parameters. It would require highly efficient sampling algorithms as well as considerably higher compute cost. Whichever approach is better remains to be seen and we leave it as a problem for future investigation. 

The most obvious targets for such analysis would be either photometric galaxy clustering or weak gravitational lensing, where the bulk modes are not directly accessible. The photometric galaxy clustering has the advantage in that while the galaxies are observed in projected, the relevant redshift range is still relatively small. However, the galaxies are non-linear tracers of the (three-dimensional!) matter fields and therefore this needs to be properly taken into account. Weak lensing, on the other hand, is a much more direct tracer of the matter fields (albeit baryonic effects and tidal alignments complicate the picture), but the weak-lensing kernel is considerably broader, spanning a significant cosmic history. This poses two problems: i) the same observed angular scales probes a range of physical scales and ii) the universe evolves considerably, so the suppression kernel $\Sigma^2$ cannot be assumed to be a single number. These issues far exceed the scope of this paper, but present and an interesting research program for the future \cite{arXiv:1203.3639,arXiv:2403.03220,arXiv:astro-ph/9509005,arXiv:0711.2521,arXiv:astro-ph/0604361,arXiv:astro-ph/0112551,arXiv:astro-ph/0509419}.
 
\section*{Acknowledgments}

KH acknowledges support from the Department of Energy  Science Undergraduate Laboratory Internships (SULI) program. Authors acknowledge useful discussions with David Alonso.

\section*{Use of AI}

The theory of Section \ref{sec:theory}, the design of the simulation suite and the analysis of Section \ref{sec:simulations} are the work of the authors. Agentic AI coding tools (OpenAI Codex GPT 5 series and Anthropic Claude Code Opus 5.5 and Fable 5.2) were used extensively for the likelihood part of this work. The differentiable forward model, the likelihood and the sampler of Section \ref{sec:likelihood} were written by these tools under the direction of AS. They were also used in the additional analysis required to the second revision of this paper and the agentic reports were used as basis for the revision. In all cases the scientific questions, the choice of the model and the interpretation were set by the authors. Authors has reviewed the code, the numerical results and the text. We take full responsibility for all claims made in this paper.

\bibliography{main}

\appendix

\section{Projections along $x$-axis}\label{app:x}
\begin{figure}
    \includegraphics[width=\linewidth]{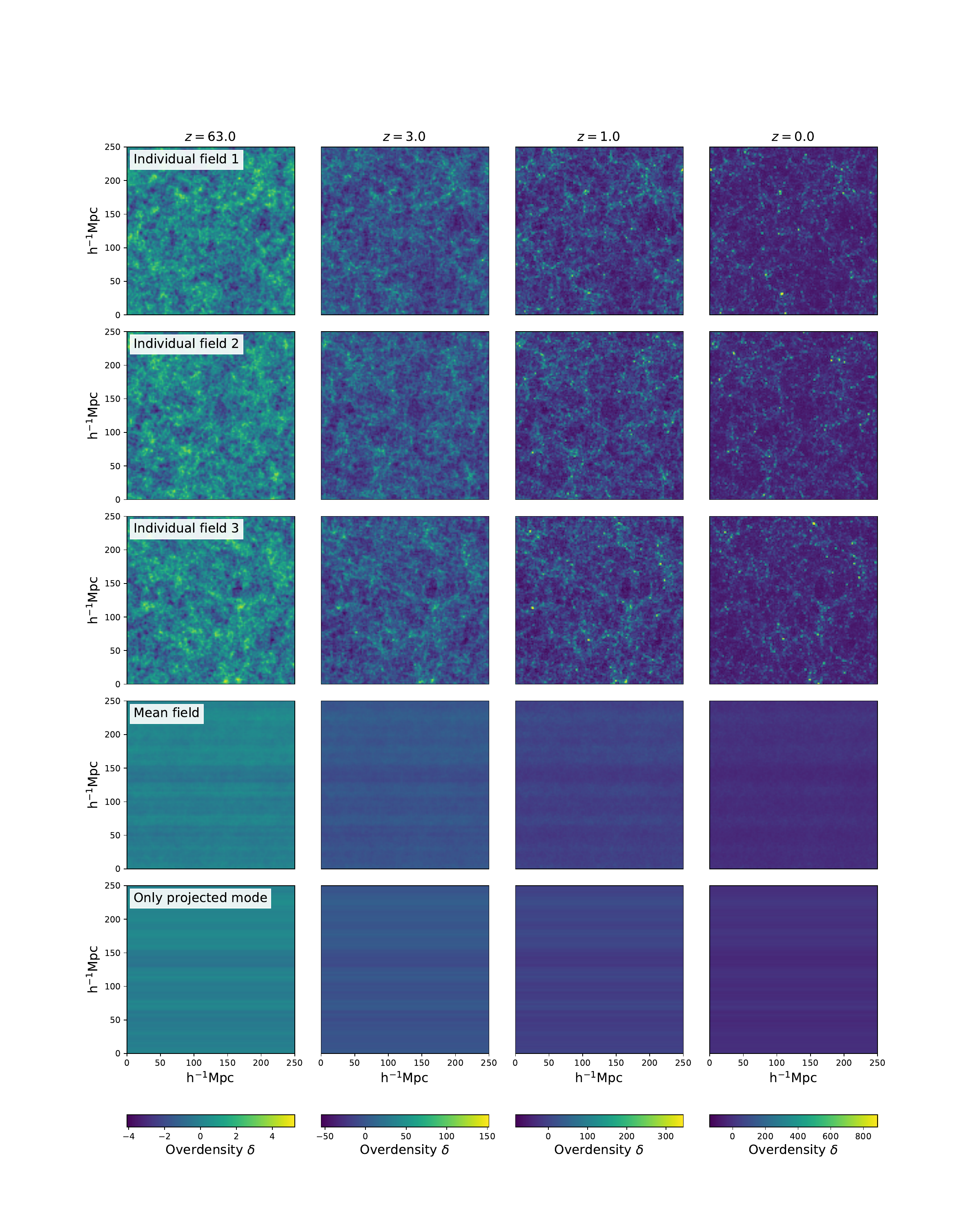}
    \caption{Same as Figure \ref{fig:projplots} but for a projection along the $x$-axis.}
    \label{fig:x-axis}
\end{figure}

In Figure \ref{fig:x-axis} we show the equivalent of Figure \ref{fig:projplots}, but for a project along the $x$-axis. As expected, the individual fields evolve into independent realizations while mean filed and OPM fields are consistent with zero (striping is due to residual sample variance in a finite set of simulations). Quantitatively, at $z=0$ the rms of the mean field along $x$ is $0.11$, against $0.26$ for the $z$-projection, and the projected-mode-only field has rms $0.10$ along $x$ versus $0.53$ along $z$. A single full realization has rms $\approx 0.76$, so the mean of $N_{\rm sim}=100$ realizations has an expected sample-variance floor of $\approx 0.76/\sqrt{100}\approx 0.08$; the measured $x$-axis rms is close to this floor, confirming that these fields are consistent with zero up to the finite-ensemble noise that also produces the visible striping.

\section{Two-dimensional LPT}\label{app:lpt}
The forward operator $\mathcal{Z}(\Delta_0,t)$ used in likelihood  is the standard
Lagrangian perturbation theory (LPT) displacement field, evaluated in two dimensions. The
linear growth factor $D(t)$ and the second-order coefficient $3/7$ --- are those of
the full \emph{three}-dimensional background cosmology (the $3/7$ being the
Einstein--de~Sitter value, an excellent approximation for $\Lambda$CDM), since the background
expansion is unaffected by the symmetry of the perturbation.

Let $\Delta_0(\mathbf{q})$ be the initial projected over-density on the Lagrangian grid
$\mathbf{q}=(q_1,q_2)$ and $\phi$ the displacement potential, $\nabla^2\phi = \Delta_0$, i.e.\
$\phi(\mathbf{k}) = -\Delta_0(\mathbf{k})/k^2$ with $k^2 = k_1^2+k_2^2$. Particles initially at
$\mathbf{q}$ are moved to
\begin{equation}
  \mathbf{x}(\mathbf{q}) = \mathbf{q} + D(t)\,\boldsymbol{\Psi}^{(1)}(\mathbf{q})
   + D^2(t)\,\boldsymbol{\Psi}^{(2)}(\mathbf{q}),
  \label{eq:lpt:x}
\end{equation}
and the evolved projected field $\mathcal{Z}(\Delta_0,t)$ is obtained by depositing the
displaced particles back onto the mesh (cloud-in-cell) and taking the over-density.

\paragraph{First order (1LPT, Zeldovich).} The first-order displacement is the gradient of the
linear potential,
\begin{equation}
  \boldsymbol{\Psi}^{(1)} = -\nabla\phi, \qquad
  \boldsymbol{\Psi}^{(1)}(\mathbf{k}) = \frac{i\mathbf{k}}{k^2}\,\Delta_0(\mathbf{k}).
  \label{eq:lpt:1}
\end{equation}
Setting $\boldsymbol{\Psi}^{(2)}=0$ in Equation \ref{eq:lpt:x} gives the 1LPT (Zeldovich) model.

\paragraph{Second order (2LPT).} The second-order displacement is sourced by the
two-dimensional determinant of the Hessian of $\phi$,
\begin{equation}
  \boldsymbol{\Psi}^{(2)} = \frac{3}{7}\,\nabla\nabla^{-2} S^{(2)}, \qquad
  S^{(2)} = \phi_{,11}\,\phi_{,22} - \phi_{,12}^2 ,
  \label{eq:lpt:2}
\end{equation}
with $\phi_{,ij}=\partial^2\phi/\partial q_i\partial q_j$ and, in Fourier space,
$\boldsymbol{\Psi}^{(2)}(\mathbf{k}) = (3/7)\,(i\mathbf{k}/k^2)\,S^{(2)}(\mathbf{k})$. In two
dimensions $S^{(2)}$ retains the single cross term $\phi_{,11}\phi_{,22}-\phi_{,12}^2$, the
direct analogue of the $\sum_{i<j}(\phi_{,ii}\phi_{,jj}-\phi_{,ij}^2)$ sum of the 3D theory.

For comparison, the Eulerian-linear (EL) model of Section \ref{sec:like:fixed} applies the
growth factor directly to the field without displacing particles,
$\mathcal{Z}_{\rm EL}=D(t)\,\Delta_0$.

\section{Differences with respect to previous versions}\label{app:changes}
This appendix is meant for the reader of an earlier version of this paper. It lists what has changed and why. The theory of Section \ref{sec:theory} and the simulations are unchanged. The changes concern the numerical values of $\Sigma^2$ and, mainly, the likelihood of Section \ref{sec:likelihood}.

\paragraph{Values of $\Sigma^2$.} The first version of this paper quoted incorrect values of $\Sigma^2$ at $z>0$ ($103$ and $264\,({\rm Mpc}/h)^2$ at $z=1$ and $3$, with measured values quoted on the same scale). The error was in the conversion to $z>0$: the scaling with the growth factor was applied in the wrong direction, i.e.\ the $z=0$ value was divided by $D^2(z)$ instead of being multiplied by it. In addition, the prefactor of $\Sigma_Z^2$ was misprinted as $1/6\pi$ instead of $1/6\pi^2$. The theory integrals of Section \ref{sec:theory} and their $z=0$ values in Table \ref{tab:num} were correct and are unchanged, and no conclusion of the paper is affected.

\paragraph{Form of the likelihood.} In the previous version $P_{\rm tot}$ was a free parameter in every $\kperp$ bin and the log-posterior contained an additional term, $|\Delta_{\rm obs}|^2/P_{\rm tot}+\ln P_{\rm tot}$, which tied it to the measured power of the data. The data enter such an expression twice and it is therefore not a posterior of any generative model. That term was there as a regulariser, because leaving $P_{\rm tot}$ free and constrained only through the residual does not work: the initial field has as many degrees of freedom as the data and on scales where $S$ is not small it can reproduce the observed modes almost exactly. The residual in a $\kperp$ bin, and with it the inferred $P_{\rm tot}$, can then be driven to zero at a finite cost in the prior, and the posterior density diverges there because of the normalisation term $\ln N$; for a scale-invariant prior on $P_{\rm tot}$ the posterior is not even normalisable. Sampling this joint posterior for realization 88 we find that the result is set by the prior on $P_{\rm tot}$ rather than by the data: for $\kperp<0.25\,h\,{\rm Mpc}^{-1}$ the inferred $P_{\rm tot}$ ranges from a few per cent of the true total power to several hundred times its value, depending on the scale and on whether the prior is scale-invariant or flat, and the chains make excursions to $P_{\rm tot}\simeq0$. The cross-correlation of the posterior mean with the truth is hardly affected, but the scatter of the posterior samples, i.e.\ the uncertainty of the reconstruction, is. Since at fixed cosmology the total power is not a free quantity, fixing it is the consistent solution and Equation \ref{eq:like:logl} is the posterior of the generative model of Section \ref{sec:like:def}.

\paragraph{1LPT versus 2LPT.} The previous version reported a markedly better reconstruction with 1LPT than with 2LPT. That was an artefact of our sampler. Modes beyond the Nyquist frequency (the corners of the two-dimensional Fourier grid, $21\%$ of the modes), which are absent from the initial conditions of the simulations and appeared neither in the prior nor in the likelihood, were left free. They acquired large amplitudes and, through the non-linear forward model, degraded the reconstruction of the physical modes, more so for 2LPT. These modes are now set to zero and the two forward models give indistinguishable results (Sections \ref{sec:like:fixed}--\ref{sec:like:ensemble}). The statement that 2LPT is preferred when the initial field is fixed to truth is also withdrawn.

\paragraph{Mask.} In the previous version the variance $N$ of the masked residual was multiplied by $f_{\rm sky}$ and the $\chi^2$ was summed over all modes. This applies only one of the two factors of the $f_{\rm sky}$ approximation: $f_{\rm sky}N$ is the correct variance of an individual mode of the masked residual, but only a fraction $f_{\rm sky}$ of these modes are independent. The data were therefore over-weighted by $1/f_{\rm sky}$ with respect to the prior. For the estimation of a power spectrum such an omission would affect only the error bars, but here it makes the posterior too narrow and the posterior mean too large wherever the residual and the signal are comparable: repeating the tests of Section \ref{sec:like:ensemble} with the old weighting, the regression coefficient of the truth on the posterior mean drops by $5\%$, $8\%$ and $11\%$ at $\kperp\simeq0.20$, $0.27$ and $0.36\,h\,{\rm Mpc}^{-1}$ (to $0.93$ and $0.92$ in the first two bins). This agrees with the factor $1-(1-f_{\rm sky})(1-r^2)$ expected for a Wiener filter in which the noise is underestimated by $f_{\rm sky}=0.84$. In addition, the damping $S$ is now applied to the predicted field before the mask rather than after it.

\paragraph{Sampling and tests.} The chains are now started from random draws from the prior rather than from a maximum a posteriori fit, the sampler is preconditioned with the prior covariance and $P_0$ and $P_{\rm tot}$ are specified per mode rather than in coarse $\kperp$ bins. The convergence diagnostics of Section \ref{sec:like:sampling} and the ensemble tests of Section \ref{sec:like:ensemble} are new, as are Figures \ref{fig:like_resid} and \ref{fig:like_ensemble} and Table \ref{tab:coverage}; all other figures of Section \ref{sec:likelihood} have been redone. The implementation uses \texttt{PyTorch} and not \texttt{jax}, as was incorrectly stated before.

\end{document}